\documentclass[a4paper,10pt]{article}
\usepackage[pdftex]{graphicx}
\PassOptionsToPackage{hyphens}{url}\usepackage{hyperref}
\hypersetup{
     colorlinks = true,
     allcolors = blue
}
\usepackage{amsmath,amsfonts,amssymb}
\usepackage{fullpage}
\usepackage{authblk}
\usepackage{setspace}
\usepackage{subcaption}
\usepackage{booktabs}
\usepackage{tabularx}

\usepackage[explicit]{titlesec}
\usepackage{sidecap}
\usepackage{pbox}

\usepackage{amsfonts}
\usepackage{amsmath}
\usepackage{multirow}
\usepackage{longtable}
\usepackage{makecell}
\usepackage{graphicx}
\usepackage{xcolor}
\usepackage{comment}
\newcommand{\deletetext}[1]{\iffalse{{\color{red}{#1}}}\fi}
\newcommand{\deletetextnew}[1]{\iffalse{{\color{red}{#1}}}\fi}
\newcommand{\newtext}[1]{{{#1}}}
\newcommand{\newtextnew}[1]{{#1}}

\usepackage{hyperref}

\date{}
\title{
The Concept of Decentralization Through Time and Disciplines: A Quantitative Exploration
}

\author[1,2,3,+]{Gabriele Di Bona}
\author[4,+]{Alberto Bracci}
\author[1]{Nicola Perra}
\author[1,5,6]{Vito Latora}
\author[4,7,8,*]{\\Andrea Baronchelli}

\affil[1]{\small School of Mathematical Sciences, Queen Mary University of London, London E1 4NS, United Kingdom}
\affil[2]{\small CNRS, GEMASS, 59 rue Pouchet, Paris F-75017, France}
\affil[3]{\small Sony Computer Science Laboratories Rome, Joint Initiative CREF-Sony, Rome I-00184, Italy}
\affil[4]{\small City, University of London, Department of Mathematics, London EC1V 0HB, UK}
\affil[5]{\small Complexity Science Hub Vienna, A-1080 Vienna, Austria}
\affil[6]{\small Dipartimento di Fisica ed Astronomia, Universit\`a di Catania and INFN, I-95123 Catania, Italy}
\affil[7]{\small The Alan Turing Institute, British Library, 96 Euston Road, London NW12DB, UK}
\affil[8]{\small UCL Centre for Blockchain Technologies, University College London, London WC1E 6BT, UK}
\affil[+]{\small Contributed equally}
\affil[*]{\small Corresponding author: abaronchelli@turing.ac.uk}

\begin{document}
\maketitle
\begin{abstract}
\textit{Decentralization} is a \deletetext{widespread }\newtext{pervasive} concept \newtext{found} across disciplines, \deletetext{such as }\newtext{including} Economics, Political Science, and Computer Science, where it is used in distinct \deletetext{but overlapping }\newtext{yet interrelated} ways. 
Here, we \newtext{develop and publicly release a general pipeline to} investigate the scholarly history of the term\deletetext{ by}\newtext{,} analysing \deletetext{$425$k }\newtext{$425\,144$} academic publications \deletetext{mentioning }\newtext{that refer to} \emph{(de)cen\-tral\-iza\-tion}. 
We find that the fraction of papers on the topic has been exponentially increasing since the 1950s\newtext{. In 2021}, 1 author in 154 \deletetext{mentioning }\newtext{mentioned} \emph{(de)cen\-tral\-iza\-tion} in the title or abstract of an article\deletetext{ in 2021}. 
\deletetext{We then cluster papers using both semantic and citation information and show that the topic has independently emerged in different fields, while  cross-disciplinary contamination started only more recently. }%
\newtext{%
Using both semantic \newtext{information} and citation \deletetext{information }\newtext{patterns}, we cluster papers in fields and characterize the knowledge flows between them.}
\deletetext{We then cluster papers and show that the topic has independently emerged in different fields, while  cross-disciplinary contamination started only more recently. }%
\newtext{Our analysis reveals that the topic has independently emerged in the different fields, with small cross-disciplinary contamination.}
\deletetext{Finally }\newtext{Moreover}, we show how Blockchain has become the most influential field about 10 years ago, while Governance dominated before the 1990s\deletetext{, and we characterize their interactions with other fields}.
\deletetext{Our }\newtext{In summary, our findings provide a quantitative assessment of the evolution of a key yet elusive concept, which has undergone cycles of rise and fall within different fields.} 
\deletetext{Furthermore, our general framework---whose code is publicly released alongside this paper--- }%
\newtext{Our pipeline offers a powerful tool to analyze the evolution of any scholarly term in the academic literature, providing insights into the interplay between collective and independent discoveries in science.}
\deletetext{
may be used to run similar analyses on \deletetext{virtually any }other \deletetext{concept }\newtext{terms} in the academic literature.
}
\end{abstract}

\section{Introduction}

``For students of recent domestic affairs it is becoming increasingly evident that `\emph{decentralization}' is a magic word". 
With these words in 1975 Herbert London opens his article ``The meaning of decentralization"~\cite{meaning_decentralization}. 
Almost 50 years later, Schneider states that \emph{decentralization} ``is called for far more than it is theorized or consistently defined"~\cite{ambition}. 
\textit{(De)centralization} (i.e., either \textit{Decentralization} or its counterpart \textit{Centralization}) has indeed become almost a buzzword, permeating not only the academic literature, but also the public discussion. The debate between centralized and decentralized contact tracing at the beginning of the COVID-19 pandemic is a clear example~\cite{contact_tracing}. However, one of the major drivers of its growth has certainly been the rise of blockchain based technologies such as cryptocurrencies, NFTs and the metaverse.~\cite{bitcoin2008,crypto_2,crypto_3,nft_paper2}. 

However, \emph{(de)cen\-tral\-iza\-tion} is not a new concept, and has different connotations across fields. In political science, it usually refers to the delegation of power to local communities with respect to a central government~\cite{political_science_1}. The concept has similar connotations when referring to educational~\cite{education_1}, fiscal~\cite{fiscal_1} and more generally governance systems. 
Other domains where the term is widely used include public health~\cite{public_health_1}, internet protocols~\cite{internet_1}, robot swarms~\cite{robot_swarm_1} and social network analysis~\cite{freeman} among others, with the last one providing one of the few quantitative definitions available thanks to Freeman in 1978.
Given this background, some questions naturally arise: have these different disciplines independently developed the concept of \emph{(de)cen\-tral\-iza\-tion}, maybe even with different meanings (i.e., a case of polysemy)? Have they influenced each other? Which fields have been most influential to the evolution of this concept? 

Here, we address these questions by studying a corpus of scientific literature indexed by the Semantic Scholar open database~\cite{semanticscholar2018}. 
First, we observe an exponentially growing interest in the topic, with an author in 154 contributing to a paper mentioning \emph{(de)cen\-tral\-iza\-tion} in its title or abstract in 2021.
Then, we map the literature on \emph{(de)cen\-tral\-iza\-tion} by focusing on the subset of relevant articles and clustering them according to their semantic and citation information. This way, we discover 
that different academic fields have separately contributed to this topic. 
We hence study how the different clusters have influenced each other, showing how much more transfer of knowledge between different academic areas is happening in recent years. Interestingly, our analysis reveals that STEM and social sciences did not influence each other. 
Finally, we focus on two 
paradigmatic examples: Governance, interpreted generally as ``the way that organizations or countries are managed at the highest level, and the systems for doing this''~\cite{governance_def}, and Blockchain, including all blockchain based technologies, from cryptocurrencies to NFTs and the metaverse.
We show how Governance is the first cluster to extensively make use of the term \emph{(de)cen\-tral\-iza\-tion}, containing the most or second most number of papers each year since its appearance in the 1950s, and playing a leading role in the transfer of knowledge to other fields until the 1990s. Blockchain  
instead has become both the most influential cluster and the most productive cluster in the past 10 years, showing three different phases in its recent history characterized by different interactions with other fields.
Overall, our results shed light on the history and evolution of the more and more important concept of \emph{(de)cen\-tral\-iza\-tion}. Furthermore, we publicly release the code of the pipeline developed in this study, so that it may be used to study and understand the evolution of other concepts through the lenses of the academic literature.

\section{The pipeline}
\label{sec:pipeline}

In this section, we briefly describe the pipeline we have set up  
and publicly released\footnote{See \url{https://github.com/alberto-bracci/decentralization}.} 
to select the data and perform the research described in this study. 
The pipeline is conceptually divided into three steps: (1) data collection, (2) clustering of the dataset using a multilayer hierarchical stochastic block model, and (3) analysis of the influence between clusters over time using knowledge flows.

\subsection*{Step 1: Data Collection}
\label{sec:data}

The first step consists in collecting the academic publications related to the concept of \emph{(de)cen\-tral\-iza\-tion}, or potentially to other concepts.
To perform a large scale analysis of the academic literature, we exploit the possibility to access the publicly available \textit{Semantic Scholar Academic Graph} (S2AG, pronounced ``stag"), which provides monthly snapshots of research papers published in all fields~\cite{semanticscholar2018}.
Launched in 2015 by the Allen Institute for Artificial Intelligence
(AI2), Semantic Scholar provides this corpus as an open access database with the specific scope of facilitating scientific analysis of academic publications. 
It contains about 203.6\deletetext{M} \newtext{million} papers (1\textsuperscript{st} Jan. 2022 snapshot), 76.4\deletetext{M} \newtext{million} authors, and 2\deletetext{B} \newtext{billion} citations. 
Moreover, this database recently incorporated the Microsoft Academic Graph (MAG)~\cite{sinha2015overview}, which was shut down at the end of 2021~\cite{microsoft_academic_shutdown}.

From this corpus we extracted the data about papers that contain the root string \deletetext{\textit{``centraliz"} or \textit{``centralis"} }\newtext{\textit{``centrali"}} in words of the title or abstract, to capture possibly all variations of words related to the concepts of \textit{Centralization} and \textit{Decentralization} (nouns, adjectives, etc.).
In this way, we also incidentally captured articles written in different languages, mainly Portuguese, French and Spanish, and also a minority of unrelated articles (e.g., biology articles involving plant species including ``centrali" in their name).
\newtext{More information on the frequency of the words containing such root string can be found in Table~S1 in  the \textit{Supplementary Material} (SM).}
The resulting dataset has \deletetext{425k }\newtext{$425\,144$} papers characterized by a series of attributes. 
Among these, of particular interest to us there is the title, the abstract, the authors, all in- and out-citations (respectively citations and references), the year of publication, and the fields of study, which were determined based on machine learning field classifiers leveraging on the existing MAG taxonomy and classification~\cite{s2_fos}. Notice, however, that some articles miss one or more of these attributes. See Table~S\deletetext{1}\newtext{2} in the \deletetext{\textit{Supplementary Material} (}SM\deletetext{)} for details on how many papers have each of these attributes.

\subsection*{Step 2: Hierarchical clustering}

In the Semantic Scholar corpus almost each paper is associated to a list of fields of study. 
However, these are high-level, as there are in fact only 19 fields of very heterogeneous sizes (see Table~S\deletetext{2}\newtext{3} in the SM for details on how many papers are classified in each field of study).
Moreover, sometimes the fields are not correctly assigned.  
In the second step of the pipeline, we hence use a multilayer hierarchical stochastic block model (hSBM)~\cite{peixoto_science_advances,peixoto_epj}, developed to find statistically significant clusters at multiple hierarchical levels for the analysis of text data with multiple data types.
Here, in fact, we consider two layers. The document layer ---where links represent citations between papers--- and the text layer ---a bipartite network between documents and the words present in their titles. 
The method naturally produces clusters of documents and topics (word clusters), incorporating the information from both layers in the process. 
Furthermore, as the name suggests, the model produces a hierarchical clustering, providing a richer structure of both article clusters and topics, which captures both small clusters and topics and how they are related to each other in a higher level structure.

We consider only the papers in our dataset that have a non-empty title and contain at least one citation or reference to another paper in the dataset ($42.7\%$ of the initial dataset), as we are interested in how the concept of \emph{(de)cen\-tral\-iza\-tion} evolved in the academic literature, and citations are the most natural proxy for how knowledge is transferred. 
We use title texts, instead of abstracts, for various reasons: firstly, because the title is more frequently available than the abstract (see Table~S\deletetext{1}\newtext{2} in the SM); secondly, because the title has the advantage of being more distilled compared to abstracts~\cite{milojevic_1}; lastly, because titles contain a significantly smaller number of words than abstracts, allowing us to obtain a text layer similar to the document layer in terms of number of edges by simply cutting out words present in less than 5 documents. 
It is indeed well known that the hSBM performs optimally when both layers have roughly the same size, otherwise the smaller layer is effectively ignored by the algorithm~\cite{peixoto_epj}.
The filtered dataset hence consists of 181\,605 documents and 15\,381 different words, summing up to 590\,215 document-to-document citation links and 1\,396\,830 document-to-word links.

\deletetext{To make sure results are robust, the algorithm is run 100 times, and the consensus partition between the 100 runs is then computed. }
\newtext{
To ensure the robustness of our results, we performed 100 iterations of the algorithm. 
Notice that the number of clusters and levels of granularity obtained is not fixed, but is automatically suggested by the algorithm.
By running the algorithm multiple times, we aimed to capture the inherent variability and uncertainty in the Monte Carlo partitioning process. 
Subsequently, the consensus partition is calculated by maximizing the overlap with all the partitions from the 100 runs.
Such consensus partition serves as a robust representation of the underlying structure within the analyzed data.
More statistics comparing the single iterations of the hSBM and the consensus partition are shown in Fig.~S2 in the SM.
}

Afterwards, keywords are assigned to each cluster to roughly represent the content and themes of the articles within them (for more details see Fig.~S\deletetext{2}\newtext{3}, Fig.~S\deletetext{3}\newtext{4} and Table~S\deletetext{3}\newtext{4} in the SM, with related section).
Keywords are chosen by looking at the most frequent words in the cluster, the most significant topics in the cluster according to the normalized mixture proportion~\cite{peixoto_epj}, as well as the first 5 papers in the cluster according to different measures (see SM Section~2.1 for more details). 

\subsection*{Step 3: Knowledge flows}

In the third step of the pipeline, we want to better understand how the different groups of documents identified by the hSBM have influenced each other throughout history.
To do so, we evaluate the knowledge flows between these groups, using article citations as proxy~\cite{knowledge_flow}. 
In particular, we compute the knowledge flow from one cluster $a$ in one year $Y_a$ to another cluster $b$ in a future year $Y_b$. 
The computation takes into account the fraction of citations towards papers in $a$ of the year $Y_a$ from papers in $b$ published in the year $Y_b$ with respect to the fraction of citations towards $a$ in $Y_a$ from all papers published in $Y_b$, as well as the overall fraction of papers of $a$ in $Y_a$. 
The citation network suffers indeed from a series of inherent biases: field size, typical number of citations in a field or a journal, typical number of references, age of the fields etc. 
This method de facto considers the number of citations with respect to a null model, resulting in a link weight which is effectively a z-score.

Mathematically, if a collection of papers is divided in a partition $\mathcal{P}$ of clusters such that different clusters do not overlap and altogether form the collection of papers, then we can define the knowledge flow units $C_{a \to b} (Y_a, Y_b)$ from papers in cluster $a \in \mathcal{P}$ published in the year $Y_a$ to papers in cluster $b \in \mathcal{P}$ in a future year $Y_b$ by counting how many citations have occurred from $b$ to $a$ in the two years, that is,
\begin{equation}
    C_{a \to b} (Y_a, Y_b) = |\{(x,\,y) : x \in a,\, Y_x=Y_a, \, y \in b,\, Y_y=Y_b \,\text{ s.t. } y \text{ cites } x\}|.
\end{equation}
As said before, we need to normalize this number with respect to a null model, so as to keep into account different sizes of clusters and different norms in citation practices. Hence, the knowledge flow $K_{a \to b} (Y_a, Y_b)$ from $a$ in year $Y_a$ to $b$ in year $Y_b$ can be computed in the following way:
\begin{equation}\label{eq:def_normalized_knowledge_flow}
    K_{a \to b} (Y_a, Y_b) = 
    \begin{cases}
    1 & \text{if} \quad \dfrac{C_{a \to b} (Y_a, Y_b)}{\sum_{c \in \mathcal{P}}C_{c \to b} (Y_a, Y_b)}\bigg/\dfrac{|x \in a : Y_x=Y_a|}{\sum_{c \in \mathcal{P}} |x \in c : Y_x=Y_a|} \geq 1 \\
    0 & \text{otherwise}
    \end{cases} 
    \ ,
\end{equation}
After the normalization against the null model, knowledge flows can be indeed treated as z-scores. Hence, in Eq.~\eqref{eq:def_normalized_knowledge_flow} we consider a knowledge flow as significant (i.e., a binary value of 1) if higher than the threshold 1, and as not significant (i.e., 0) otherwise.

Therefore, we obtain a binary value for each pair of clusters and each pair of years.
In other words, the collection of (knowledge flow) links between all pairs of clusters and years generates a temporal network of clusters, which we aggregate in different ways to facilitate the following analysis.
In particular, we consider the average knowledge flow $K_{a \to b}(Y_a)$ from a cluster $a$ to another $b$ from a specific year $Y_a$ as the average of the knowledge flows $K_{a \to b} (Y_a, Y_b)$ from cluster $a$ to $b$ from year $Y_a$ to all years $Y_b>Y_a$, taking into account only years $Y_b$ where there is at least one publication in $b$. Formally, this reads:
\begin{equation}
    \label{eq:average_kf_to_future}
    K_{a \to b} (Y_a) = K_{a \to b} (Y_a, \bullet) = \left\langle 
    K_{a \to b} (Y_a, Y_b)
    \right\rangle_{\displaystyle\{Y_b > Y_a : \, \exists x \in b \text{ s.t. } Y_x=Y_b\}} \ .
\end{equation}
This value represents, on a scale from 0 to 1, how much publications in cluster $a$ in year $Y_a$ have influenced the future of cluster $b$. 
Analogously, we define the average knowledge flow $K_{a \to b}(T)$ from cluster $a$ to cluster $b$ from a period of time $T$ to the future by averaging $K_{a \to b} (Y_a)$ over all years $Y_a$ in $T$ in which there is at least one publication in $a$% as $K_{a \to b}(T) = \langle K_{a \to b}(Y) \rangle_{Y \in T}$
, that is, 
\begin{equation}
    K_{a \to b} (T) = \left\langle 
    K_{a \to b} (Y_a)
    \right\rangle_{\displaystyle\{Y_a \in T : \, \exists x \in a \text{ s.t. } Y_x=Y_a\}} \ .
\end{equation}

We can also measure the average influence in terms of knowledge flows from a cluster to all other clusters and vice-versa, as well as the average knowledge flow among all clusters, respectively as follows:  
\begin{equation}
\begin{split}
    \label{eq:average_kf_cluster_specific_year}
    &K_{a \to \bullet} (Y) = \left\langle 
    K_{a \to b} (Y)
    \right\rangle_{b} \  , \ \\
    &K_{\bullet \to a} (Y) = \left\langle 
    K_{b \to a} (Y)
    \right\rangle_{b} \  , \ \\
    &K_{\bullet \to \bullet} (Y) = \left\langle 
    K_{a \to b} (Y)
    \right\rangle_{a,b} 
    \ .
\end{split}
\end{equation}
Here, $K_{a \to \bullet} (Y)$ refers to the average influence from papers in cluster $a$ published in year $Y$ towards all clusters in the future. On the opposite, $K_{\bullet \to a} (Y)$ refers to the average influence of the papers in all clusters in the year $Y$ towards the future of cluster $a$. Finally, $K_{\bullet \to \bullet} (Y)$ refers to the average influence (towards the future)
of all papers published in year $Y$. 

\section{Results}

\subsection{The decentralized evolution of (de)centralization}
\begin{figure}[ht!]
    \centering
    \includegraphics[width=0.95\linewidth]{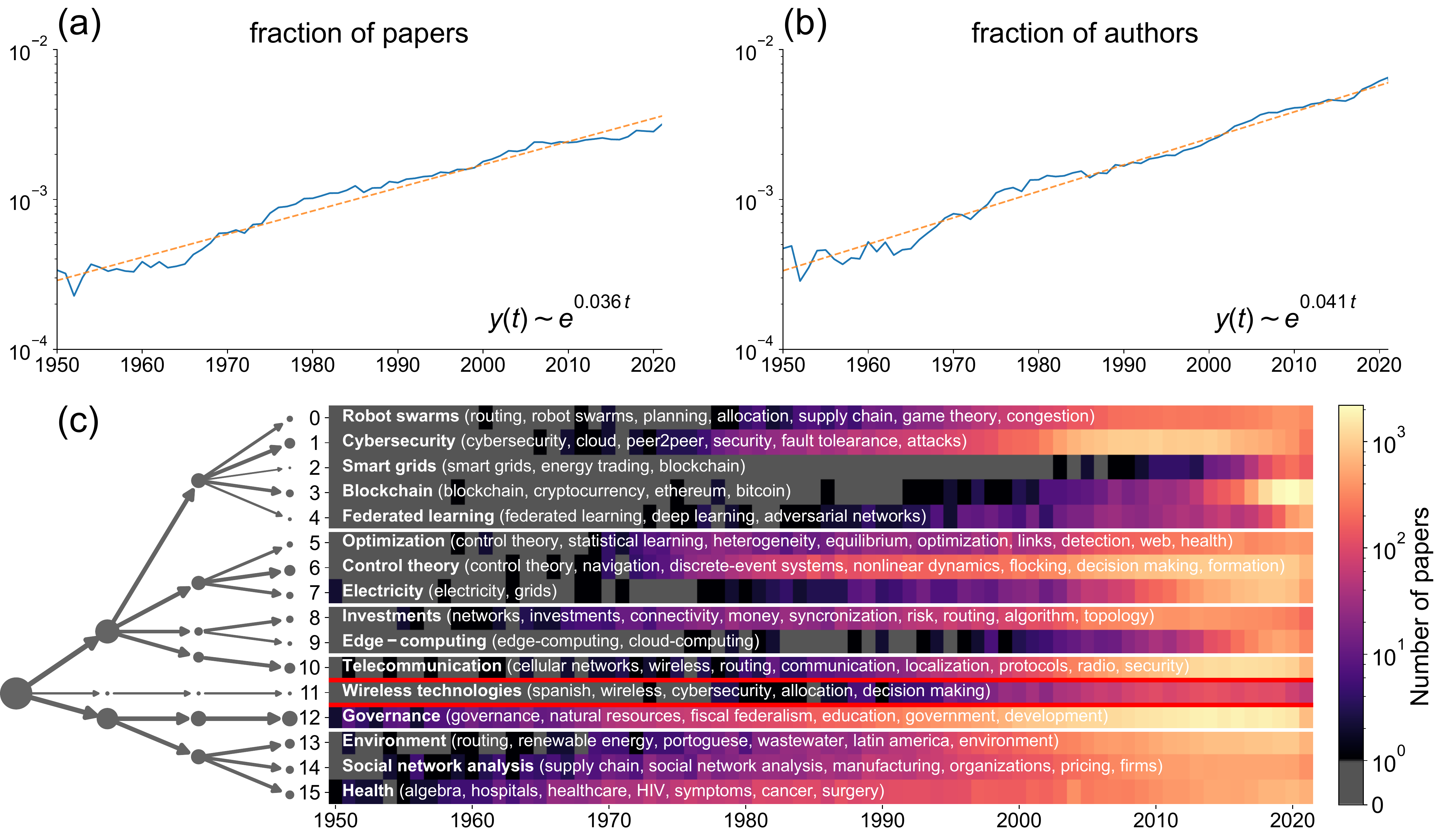}
    \caption{
    \textbf{The rising interest of academic literature towards \emph{(de)cen\-tral\-iza\-tion}.}
    \textbf{(a)} Fraction of papers mentioning \emph{(de)cen\-tral\-iza\-tion} in the Semantic Scholar corpus.
    \textbf{(b)} Fraction of authors mentioning \emph{(de)cen\-tral\-iza\-tion} in the Semantic Scholar corpus.
    Both fractions have been steadily increasing since the 1950s, showing growing interest in the topic.
    \textbf{(c)} Number of papers in the clusters at the 3\textsuperscript{rd} level of the hierarchy in each year. 
    Clusters are ordered respecting the hierarchical network on the left, in which node and link sizes are proportional to the total number of papers in the related cluster. In the heatmap, white lines individuate clusters belonging to the same cluster at the 2\textsuperscript{nd} level of the hierarchy, while red lines divide different clusters at the 1\textsuperscript{st} level. The representative keyword of the clusters at level 3 is reported in bold in the respective rows, while all the specific keywords identified at the 4\textsuperscript{th} hierarchical level are shown within brackets. Clusters with less than 500 papers in total are not shown in the figure. 
    }
    \label{fig:time_evolution_interest}
\end{figure}

We start by analysing the number of papers mentioning \emph{(de)cen\-tral\-iza\-tion} over the years (see \textit{The pipeline} section for more details), comparing it to the total number of academic outputs (papers, books etc.) produced in time, which is known to be exponentially increasing~\cite{exponential_growth}.
As shown in Fig.~\ref{fig:time_evolution_interest}(a), the fraction of papers mentioning \emph{(de)cen\-tral\-iza\-tion} has been exponentially increasing in time since the 1950s, rising to around one paper every 315 in 2021.
The growing interest in this topic is also reflected by the increasing number of authors involved in such academic research. 
Indeed, as shown in Fig.~\ref{fig:time_evolution_interest}(b), the fraction of authors producing such research has risen exponentially by more than one order of magnitude, with almost one academic every 154 writing a paper mentioning the topic in 2021. 
This growth is also seen in terms of raw number of publications and authors, as shown in Fig.~S1 in the SM, where we compare these numbers for the S2AG corpus and the \emph{(de)cen\-tral\-iza\-tion} dataset and find a stronger exponential rise for the latter.
Notice that for both papers and authors there are some periods with a higher or lower increase in the fraction, showing spikes of interest at particular times. For example, in Fig.~\ref{fig:time_evolution_interest}(a,b) we can see that between the late 1970s and the 1980s the growth rate was faster than the overall exponential fit.

In order to understand what has characterized the origins and evolution of the topic, we set to identify topics and clusters of papers in the dataset by using the hSBM algorithm~\cite{peixoto_science_advances, peixoto_epj} described in \emph{The pipeline} section. 
In the following analysis, we focus our attention only to years after 1950. 
Before this date there are only around 100 papers in our \emph{(de)cen\-tral\-iza\-tion} dataset. The very first is a political science one from 1851 on local self-governments versus centralized governments~\cite{first_paper}.
Among the others in this period, apart from around 50 papers that relate to \emph{(de)cen\-tral\-iza\-tion} in governments, organizations and states, we have detected 30 papers that are actually false positives of the selection process. Considering also how, in general, digitalization issues may have contributed to the small number of papers before 1950, the reliability and coverage of the first 99 years of the data are unclear, and we opted to exclude them from the analysis.

\deletetext{The hSBM algorithm identified 7 hierarchical levels of clusters of documents. }
\newtext{
The consensus partition, obtained by collecting the outcomes of 100 runs of the hSBM algorithm, consists of 7 hierarchical levels. 
}
On the left of Fig.~\ref{fig:time_evolution_interest}(c), we draw this hierarchy only until the 3\textsuperscript{rd} level (starting from the common root at level 0) for visualization clarity. 
\newtext{More information on the number of clusters in each level can be found in Sec.~S2.1 in the SM, where we also conduct a comparative analysis of the consensus partition with the individual runs of the hSBM algorithm.}

On the right of Fig.~\ref{fig:time_evolution_interest}(c), we also show the heatmap of the number of papers in time for each cluster at the 3\textsuperscript{rd} level, for a total of 16 different clusters after excluding other 5 clusters with less than 500 papers not included in this analysis. 
The keywords shown in the heatmap have been manually assigned to roughly characterize each cluster. 
In particular, the keywords between parentheses have been chosen amongst the most frequent and most significant in the clusters at the 4\textsuperscript{th} level, while the most representative keyword at the 3\textsuperscript{rd} level has been chosen and printed in bold. 
For details on how they were assigned see \textit{The pipeline} section and Sec.~S2\newtext{.2} in the SM.

In the following, we refer to a cluster at the 3\textsuperscript{rd} level by its representative keyword (capitalized).
As shown by the hierarchy and by the horizontal red lines in the heatmap, clusters are divided in three main branches. Looking at the two biggest branches, we can see a clear division between more STEM oriented documents (top branch) and those in 
Political sciences, Social sciences, as well as Medicine (bottom branch). 
Notably, a third smaller branch appears isolated, including papers at the intersection of the other two, mostly about Wireless technologies and their applications.
Going into more details, in the STEM branch we notice how Cybersecurity, Control theory and Telecommunication (clusters 1, 6, and 10 respectively) are the ones producing most publications, with Blockchain (cluster 3) becoming the most relevant in the last 5 years in terms of number of papers published per year. 
On the other branch, Governance (cluster 12), including works in Political science, Education and Fiscal federalism, is the most relevant cluster, while Environment, Social network analysis and Health clusters (respectively clusters 13, 14, and 15) have produced a smaller number of papers.
Furthermore, see Fig.~S\deletetext{4}\newtext{5} and Fig.~S\deletetext{5}\newtext{6} in the SM respectively for a similar plot done at the 4\textsuperscript{th} level and for a bipartite hierarchical network showing how clusters are represented in the various topics.

\deletetext{As said, }\newtext{In Fig.~\ref{fig:time_evolution_interest}(c) we have shown} \deletetext{we show }how the number of papers in each of these clusters has evolved over time. 
Looking more into details of the early history of \textit{(de)centralization}, the first papers 
adopting the term have all been in the Social sciences branch, most importantly the Governance cluster, followed by Social network analysis and Health.
In the 1950s, indeed, there are 58 papers in Governance, which represents the first cluster to adopt the term and use it extensively. Some of these articles refer, among other things, to democracy as a form of centralized decision-making system~\cite{1950_2}. Other clusters with more than 10 papers refer to Social network analysis and Health, as seen for example in Kaufman's \deletetext{"}\newtext{``}Toward an interactional conception of community"~\cite{1950_1}. Here, \textit{centralization} is depicted as a force gradually destroying the concept of community as a social unit. Notably, most of these papers have no citations from other articles in the full corpus, with only some citations within the cluster of governance.
In the 1960s, the largest growth is found again in the Governance and Health clusters, both reaching around 150 papers each in the decade. An important example of the former is that of Bachrach et al.~\cite{1960_1}, where they highlight how different disciplines (i.e., social and political sciences) reach completely opposite conclusions about the \emph{(de)cen\-tral\-iza\-tion} of power. In the same decade \emph{(de)cen\-tral\-iza\-tion} also appears in other relevant clusters, namely Social network analysis (50 papers) and Investments (29 papers), with a significant number of citations in both directions between them.
The term is picked up from the STEM branch only later in the 1970s, especially through works in Control theory and Optimization~\cite{SILJAK19781849}, coming significantly to a popular domain as Cybersecurity only in the 1980s.

\medskip

\begin{figure}[ht!]
    \centering
    \includegraphics[width=0.95\linewidth]{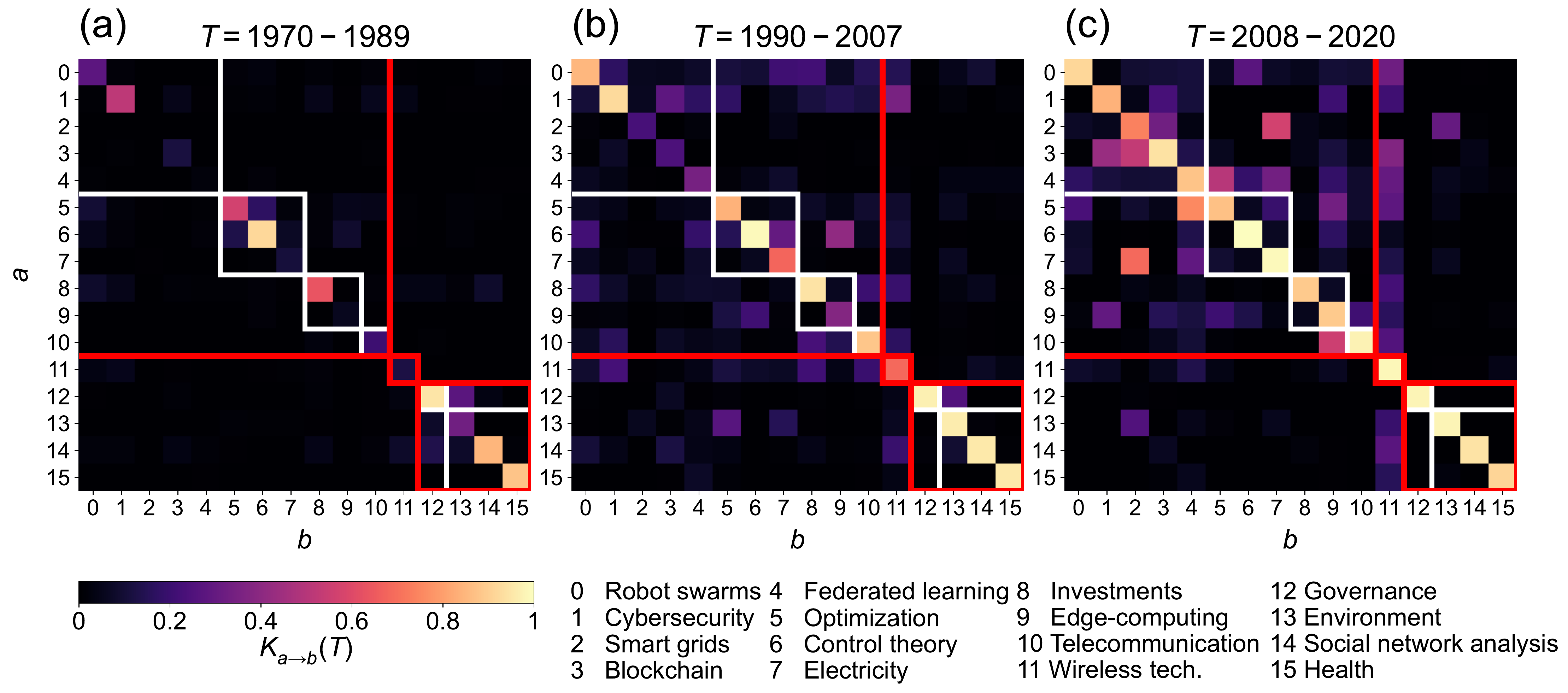}
    \caption{\textbf{Temporal evolution of the influence between clusters.} Average knowledge flows $K_{a \to b}(T)$ from each cluster $a$ to each cluster $b$ at level 3 in the period $T$ = 1970-1989 \textbf{(a)}, $T$ = 1990-2007 \textbf{(b)} and $T$ = 2008-2020 \textbf{(c)}, represented by the different colors according to the colorbar in the bottom left part of the figure. A representative keyword for each cluster is reported on the bottom right part of the figure.
    White lines denote clusters belonging to the same 2\textsuperscript{nd} level cluster, whereas red lines mark different branches at the 1\textsuperscript{st} level.
    In the first period, little to no communication is happening between different clusters. 
    In recent years, more communication happens inside the same 2\textsuperscript{nd} level cluster, and towards the middle branch (cluster 11). 
    However, little communication happens between the two other branches, roughly representing the STEM and social sciences communities respectively.
    }
    \label{fig:knowledge_flows_all}
\end{figure}

We have seen how different domains have picked up the concept of \emph{(de)cen\-tral\-iza\-tion} at different times.
It is therefore natural to ask whether they developed such uses separately, or they influenced each other in some way. 
The hierarchical clustering partially answers this question, as it gives a degree of separation between domains based on citation and semantic information.
However, significant information is still present in the citations between papers of different clusters.
We exploit this by computing knowledge flows~\cite{knowledge_flow}, whose aim is to quantify the transfer of knowledge given by the citations between groups of papers through a comparison with a null model (for more details see \textit{The pipeline} Section).
\deletetext{Here, we }\newtext{We thus} study the average knowledge flow $K_{a \to b}(T)$ from papers in a cluster $a$, at level 3 of the hierarchy, in a period of time $T$ to future publications in another cluster $b$, represented by a number between 0 and 1 showing how significant this influence has been.

\deletetext{In particular }\newtext{Here}, in Fig.~\ref{fig:knowledge_flows_all} we consider three different periods of time $T$: 1970-1989, 1990-2007 and 2008-2020. 
Similarly to what we will do in the next figures, we have excluded the year 2021 as a source of knowledge flow, because our dataset ends at the end of 2021, thus meaning that we cannot evaluate knowledge flows from papers of 2021 to future years.
In the figure, all clusters are ordered as in Fig.~\ref{fig:time_evolution_interest}(c), with the representative keywords shown in the legend below. 
For each period $T$, the color of the cell of row $a$ and column $b$ of the heatmap refers to the average knowledge flow $K_{a \to b}(T)$ from papers in cluster $a$ in that period of time to future papers in cluster $b$, according to the colormap shown below.

\deletetext{As seen }\newtext{Starting from the first period} 
in Fig.~\ref{fig:knowledge_flows_all}(a), between 1970 and 1989 clusters have little to no influence on the future of the other ones. We have previously seen how in these years the use of \emph{(de)cen\-tral\-iza\-tion} started to rise across some domains, mostly being Governance, Control theory, Social network analysis, Health, Cybersecurity, and Investments. 
However, apart from Governance and Control theory (clusters 12 and 6), these clusters have low knowledge flow even to themselves, meaning that the use of \emph{(de)cen\-tral\-iza\-tion} was only relegated to sporadic and not so influential papers in the literature. 
This also confirms that the topic has appeared independently at this early stage.

\deletetext{In }\newtext{Then, looking at the period from 1990 and 2007 in} Fig.~\ref{fig:knowledge_flows_all}(b)\deletetext{ instead}, we can see how much more transfer of knowledge has occurred between clusters\deletetext{ from 1990 and 2007}. 
As shown in Fig.~\ref{fig:knowledge_flows_all}(c), this trend is even more pronounced in \deletetext{recent years }\newtext{the last and more recent period}, \deletetext{which notably }\newtext{whose start} coincides with the creation and rise of blockchain technologies.
Interestingly, these transfers reflect the structure of the hierarchy and denote significant differences between the high-level domains. 
The STEM branch (made of the clusters 0 to 10) shows clear communication between clusters belonging to the same group both at the 2\textsuperscript{nd} and 1\textsuperscript{st} level (respectively within white and red lines), whereas the right bottom branch shows almost no communication with the other domains, especially after 2008. The only significant knowledge flow from this branch in the middle period goes from Environment (cluster 13) to Optimization (cluster 5), while in the last period this is only relegated between Environment and Smart grids (cluster 2).
The middle branch instead shows clear influence from the other two, and little influence towards them, especially in the last period.

\deletetext{Moreover }\newtext{Finally}, notice how the highest knowledge flows between different clusters in Fig.~\ref{fig:knowledge_flows_all}(b) are from those that, in the first period, were starting to be more influential\deletetext{, while }\newtext{.
Similarly} in the last period \deletetext{it is relegated }\newtext{the clusters with highest influence on other ones are} mostly \deletetext{to just STEM clusters }\newtext{within the STEM branch}. 
\deletetext{Put together }\newtext{Overall}, the heatmaps show a clear decentralized birth of the concept of \emph{(de)cen\-tral\-iza\-tion}, appearing in different fields and domains with little to no communication between each other. Instead, in recent years, we find a more coordinated evolution, even though still sectorial in some cases, and mainly lead by STEM related clusters.

\subsection{The case of Governance and Blockchain}

\begin{figure}[ht!]
    \centering
    \includegraphics[width=0.95\linewidth]{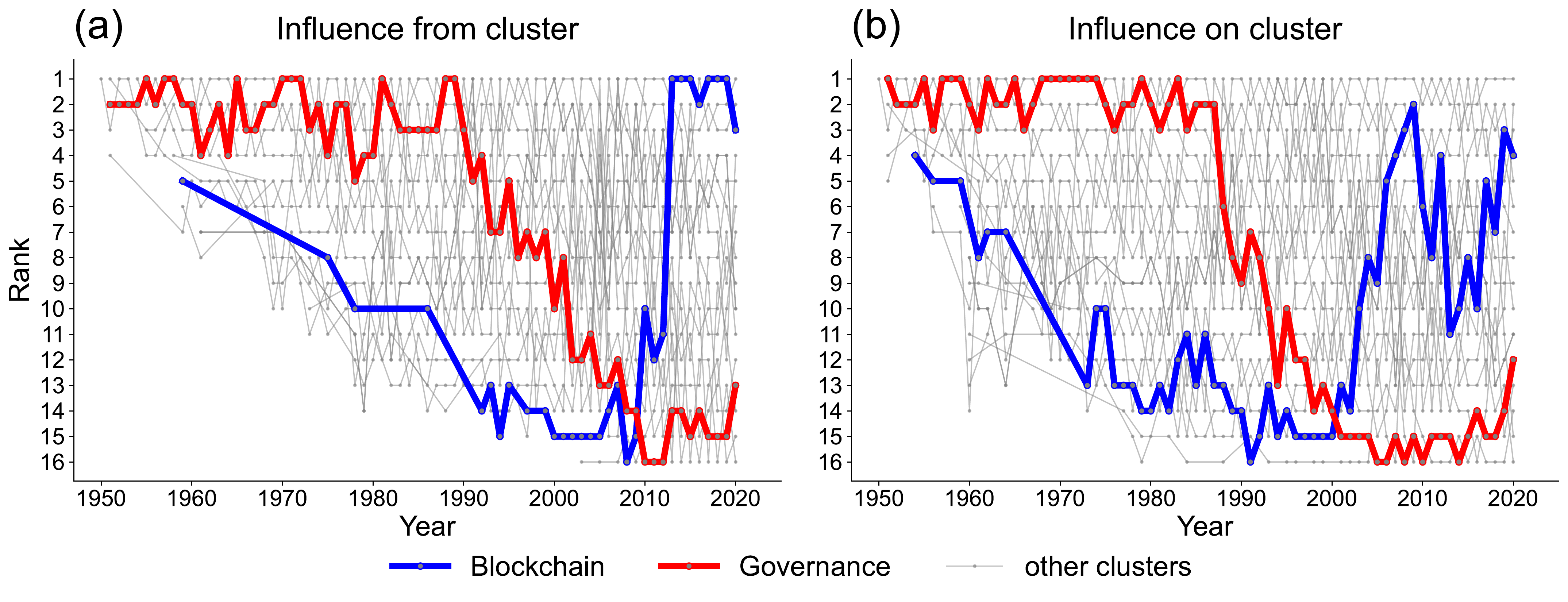}
    \caption{\textbf{Ranking the influence of Blockchain and Governance in the  \emph{(de)cen\-tral\-iza\-tion} literature.}
    \textbf{(a)} Ranking in time of the influence coming from a cluster in the 3\textsuperscript{rd} hierarchical level, computed on the average knowledge flow $K_{a \to \bullet} (Y)$ from papers in cluster $a$ published in the year $Y$ towards all other future papers.
    \textbf{(b)} Ranking in time of the influence to a cluster in the 3\textsuperscript{rd} hierarchical level, computed on the average knowledge flow $K_{\bullet \to a} (Y)$ from papers published in the year $Y$ towards future papers in cluster $a$.
    The blockchain cluster, highlighted in blue, has become a central actor in the recent literature on \emph{(de)cen\-tral\-iza\-tion}, supplanting the governance cluster, highlighted in red. 
    }
    \label{fig:rank}
\end{figure}

Having analyzed the concept of \emph{(de)cen\-tral\-iza\-tion} in the general academic landscape, we now focus on two of the most important clusters in the history of this topic: Governance and Blockchain. 
As shown in Fig.~\ref{fig:time_evolution_interest}(c) and in Fig.~S\deletetext{6}\newtext{7} in the SM, these two clusters are among the biggest across time in terms of number of papers. 
The Governance cluster has always been first or second with respect to the other clusters at the 3\textsuperscript{rd} level, while Blockchain was barely present before 2008, the year of the bitcoin white paper~\cite{bitcoin2008}. 
After that, Blockchain gradually increased in size and had an exponential explosion after 2015, coincidentally with the increasing hype around the technology and its applications, in particular bitcoin and ethereum~\cite{hype,mclean2016demystifying,ametrano2016bitcoin}. Finally, it has become the most productive cluster since 2019, surpassing governance.

To better understand their role in the evolution of the literature 
on \emph{(de)cen\-tral\-iza\-tion}, we consider the average knowledge flows between clusters for each year, that is looking at $K_{a \to \bullet} (Y)$, $K_{\bullet \to a} (Y)$, and $K_{\bullet \to \bullet} (Y)$, defined in Eq.~\eqref{eq:average_kf_cluster_specific_year}. 
Therefore, in Fig.~\ref{fig:rank} we rank clusters year by year using $K_{a \to \bullet} (Y)$ in (a) and $K_{\bullet \to a} (Y)$ in (b), i.e., looking at how much the papers of a cluster $a$ in a year $Y$ have influenced, on average, the future of all other clusters (a), or, vice versa, how much all clusters have influenced the future of $a$ (b). 
From these plots we can see how, on the one hand, Governance has been in the top ranks until the late 1980s, both as a source and target of knowledge flows. However, in the early 1990s it started to decrease in importance, reaching the bottom ranks in the 2000s, despite being the first cluster in terms of number of papers each of these years.
On the other hand, in Fig.~\ref{fig:rank}(a) we notice that the rise of Blockchain started only in 2010, being almost always outside of the \emph{(de)cen\-tral\-iza\-tion} literature discussion until this point.
Then, very sharply, Blockchain becomes the first cluster in terms of influence towards other clusters in 2013, maintaining its position in the following years. Hence, the literature on Blockchain has been key in the development of the \emph{(de)cen\-tral\-iza\-tion} discussion in the most recent years. Moreover, looking at Blockchain in Fig.~\ref{fig:rank}(b), papers of other clusters before early 2000s have had almost no impact on the scientific future of Blockchain. 
Interestingly, it has received a lot of influence from publications between 2006 and 2012, that is about when the blockchain and bitcoin originated~\cite{bitcoin2008}, as well as after 2017, mostly due to the increasing amount of applications using blockchain in the most diverse contexts in recent years.
Finally, notice the loss of influence on Blockchain from papers between 2013 and 2016.

These results are corroborated by the time evolution of the average knowledge flow compared to the overall average $K_{\bullet \to \bullet} (Y)$. Indeed, in Fig.~S\deletetext{7}\newtext{8} in the SM we show how Governance has been increasingly important in influencing other clusters until the 1980s, while since the 1990s it has had a lower average knowledge flow than the average among all clusters. Similarly to what shown by the ranks, after 2013 Blockchain starts to have a much higher influence towards the other clusters compared to the average. 
\newtext{
    Moreover, in Fig S9 in the SM we compare the average knowledge flow within the same cluster and towards other clusters, isolating the clusters Governance and Blockchain from the rest of the clusters.
    We find that on the one hand Governance has maintained a very high knowledge flow to future papers in the same cluster throughout the years. On the other hand, starting from 1990s, the exchange in knowledge flow towards and from other clusters has decreased.
    Blockchain, instead, has received a higher than average knowledge flow from other clusters in the 2000s, starting to provide maximum knowledge flow to itself and more than average to other clusters in the next and most recent decade.
}

\medskip

\deletetext{W}\newtext{So far w}e have seen how \deletetext{influential }Governance has been \newtext{influential} in the early literature about \emph{(de)cen\-tral\-iza\-tion}, and how Blockchain has risen in recent years as the most important influential cluster, contributing in terms of knowledge flow towards other branches of literature.
It is therefore a natural next step to investigate \deletetext{in more details }\newtext{with higher granularity} which clusters in particular have influenced or have been influenced by Governance first and Blockchain then, and see how these interactions have changed over time.
We start this analysis from the more recent case of Blockchain. This cluster started to appear only around 2008 with the bitcoin white paper~\cite{bitcoin2008}. Moreover, we notice a decrease in the influence on this cluster in mid 2010s. We therefore divide the 2008--2020 time span in three parts, following the blockchain history: 2008--2014, representing the origin of blockchain applications before the advent of ethereum; 2015--2018, when the field got more recognition thanks to ethereum and bitcoin; and the final 2019--2020 period, in which we have seen the explosion of academic literature production and the widespread success of multiple applications such as DeFi, NFTs and the metaverse.

\begin{figure}[ht!]
    \centering
    \includegraphics[width=0.95\linewidth]{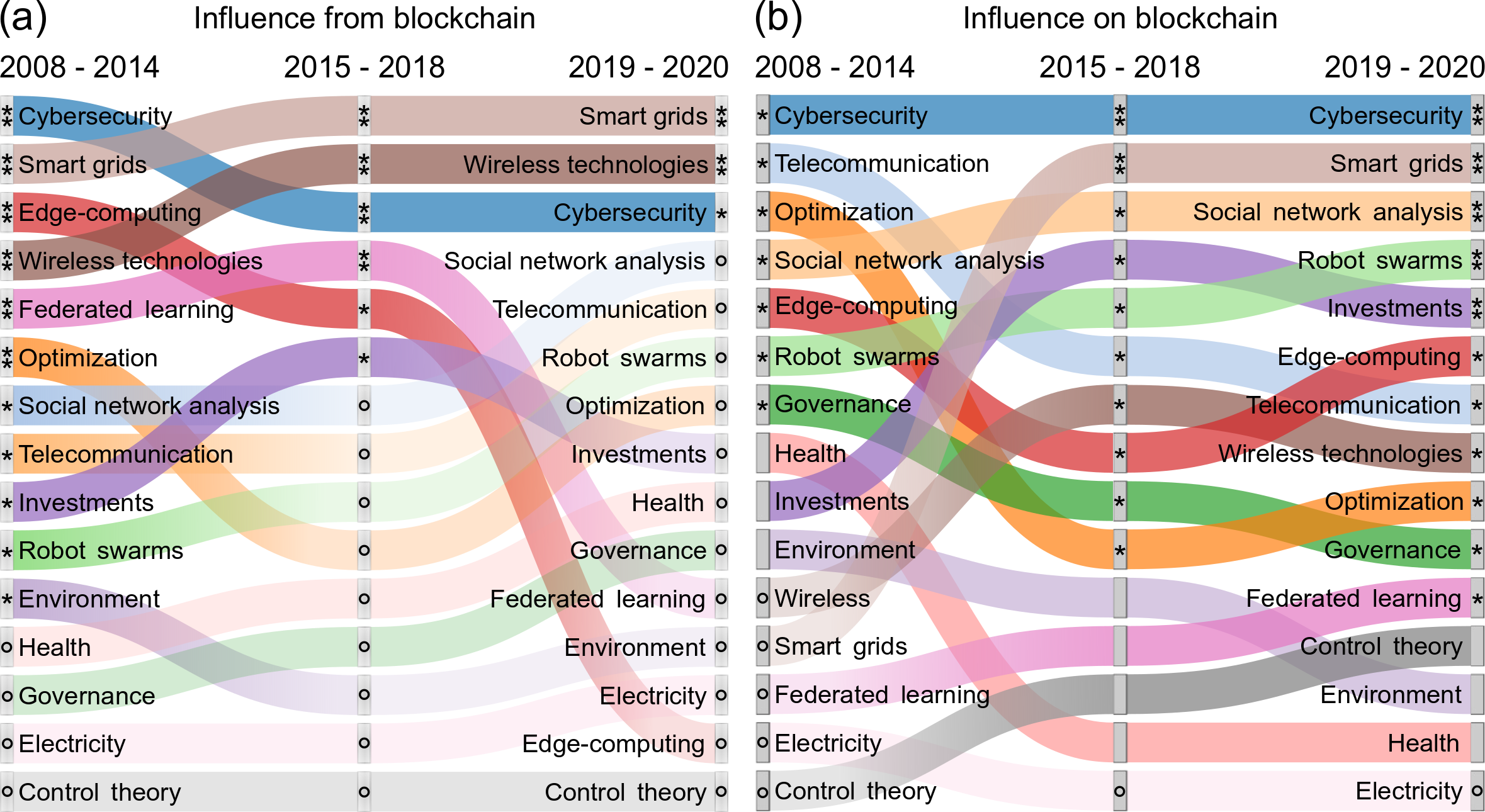}
    \caption{
    \textbf{Influences between Blockchain and the other clusters on \emph{(de)cen\-tral\-iza\-tion}.} 
    \textbf{(a)} Change in the ranking of the most influenced clusters 
    by Blockchain between its early period (2008-2014), its middle period (2015-2018), and its late period (2019-2020), calculated using the average knowledge flows $K_{a \to b} (T)$, where $T$ is the selected period, and $a$ is fixed to be Blockchain.
    \textbf{(b)} Change in the ranking of 
    the clusters having most influenced 
    the Blockchain literature (same periods 
    as in the previous panel). 
%    between its early period (2008-2014), its middle period (2015-2018), and its late period (2019-2020), calculate using the average knowledge flows $K_{a \to b} (T)$, where $T$ is the selected period, and $b$ is fixed to blockchain.
    In both cases, if $K_{a \to b} (T) = 0$, we print a circle in the corresponding gray node and use a lighter color in the respective link. Moreover, we print a star when $0.01 < K_{a \to b} (T) \leq 0.1$, and two stars when $K_{a \to b} (T) > 0.1$.
    }
    \label{fig:rank_sankey}
\end{figure}

In Fig.~\ref{fig:rank_sankey} we plot, in a decreasing order, which clusters have been most influenced by (a) and have most influenced (b) Blockchain during the three periods. To this end, we use a Sankey diagram, showing how the overall picture has changed in the three different phases.
The plot is done using the average knowledge flows $K_{a \to b} (T)$, where $T$ is the selected period, while $a$ and $b$ are fixed to Blockchain in Fig.~\ref{fig:rank_sankey}(a) and Fig.~\ref{fig:rank_sankey}(b) respectively.

\deletetext{We can see important differences across the three periods. }%
\newtext{Firstly, in Fig.~\ref{fig:rank_sankey}(a) we analyze the influence of Blockchain on the these other clusters.}
\deletetext{First, as shown by Fig.~\ref{fig:rank_sankey}(a), }%
\deletetext{t}\newtext{T}he early literature of Blockchain has had a \deletetext{big }\newtext{strong} impact on most of the clusters. As a matter of fact, there are only a few cases where the average knowledge flow from Blockchain to another cluster is zero, shown by a circle in the respective node and a lighter color in the corresponding link.
We also notice that Cybersecurity, Smart grids, Edge-computing, Wireless technologies, and Federated learning have a very significant average knowledge flow from Blockchain, i.e., $K_{a \to b} (T) > 0.1$, shown by the double stars, while other clusters with $0.01 < K_{a \to b} (T) \leq 0.1$ are represented with only one stars. 
Notice how Blockchain has continued to have a big impact on these mentioned clusters\deletetext{
. In particular, papers of Blockchain in the last period have had a significant impact on the future of only}\newtext{, in particular} Smart grids and Wireless technologies, as well as \deletetext{of } Cybersecurity to a lesser extent. 
\newtext{Looking altogether at the three periods, notice how Cybersecurity and Edge-computing have lost influence from Blockchain over time.}
\deletetext{On the contrary, }\newtext{Moreover, in the last period} there is no significant knowledge flow to all other clusters, which is peculiar if we consider that, for example, Federated learning and Edge-computing received a very significant knowledge flow in the previous years. 
We argue that \deletetext{this }\newtext{the recent} decrease in knowledge flow is mostly due to the time needed for a paper to attract citations, especially outside its own cluster.
\newtext{Additionally, we find that some clusters, such as Health, Electricity, Control theory and Governance, have received no significant influence from Blockchain in all these years, even if, Governance, for instance, has been second only to Blockchain in terms of number of papers.}

\deletetext{Looking altogether at the three periods, notice how Cybersecurity and Edge-computing have lost influence from Blockchain over time, while Smart grids and Wireless technologies have become more reliant on Blockchain with respect to the other clusters. Moreover, we find that some clusters, such as Health, Electricity, Control theory and Governance, have received no significant influence from Blockchain in all these years, even if, Governance, for instance, has been second only to Blockchain in terms of number of papers.
}%
\deletetext{When looking at Fig.~\ref{fig:rank_sankey}(b), we can see how over the years, more and more clusters have had a strong and significant impact on the future Blockchain literature. }%
\newtext{Secondly, in Fig.~\ref{fig:rank_sankey}(b) we investigate the impact of the different clusters on Blockchain's literature.}
\deletetext{In particular, }Cybersecurity, which has been one of the clusters that grew the most among all STEM clusters from the 1980s to the 2010s, has been stably the most influential cluster on Blockchain. 
The other top positions have instead changed from the first period considered, with Smart grids, which did not even have any influence on Blockchain at first, and Social network analysis becoming the \deletetext{next }most important clusters \newtext{after Cybersecurity}. 
Notice also how Robot swarms and Investments have experienced an increase in knowledge flow towards Blockchain, while the opposite has happened for Telecommunication, Optimization, Governance and Health.

Comparing the two plots in Fig.~\ref{fig:rank_sankey}, we find examples of only unidirectional influences between Blockchain and the other clusters. The cluster of Social network analysis, third in position since 2015 to influence Blockchain, has not been influenced by it during the same period, which is also the case of Robot swarms and Governance. A similar situation is found for Wireless technologies, that has been strongly influenced by Blockchain over time, but only in recent years it has had a small impact on it. 

\deletetext{W}\newtext{Finally, w}e have conducted a similar analysis on the Governance cluster in Fig.~S\deletetext{8}\newtext{10} in the SM. In this case, we consider three different periods of times: 1950--1980, that is the early stage when it was the most important cluster overall; 1981-1990, when the amount of knowledge flow from Governance stopped to increase, still remaining among the top in terms of ranking; and 1991--2000, in which its role diminished and got surpassed by almost all other clusters by the end of the period.
We do not find many noticeable differences between the first two periods. Most clusters have no significant knowledge flow from and to Governance, showing how \emph{(de)cen\-tral\-iza\-tion} developed independently in this cluster at first. Differently from Blockchain, the top clusters to have interactions with governance are Environment, Social network analysis and Investments. Wireless technologies, Blockchain and Robot swarms have also been influenced by Governance, but not vice-versa, apart from the sporadic case of Wireless technologies in the middle period. We can also see that the influence from Governance has increased over time on clusters like Blockchain, Optimization and Robot swarms, showing how the last years of the last century have been important milestones for the future of these clusters. 

\section{Discussion}

In this paper \deletetext{we have analyzed }\newtext{we have developed and presented a framework that allowed us to quantitatively investigate}  
how different topics have risen in the \emph{(de)cen\-tral\-iza\-tion} literature and have influenced it.
By exploiting the S2AG corpus, we have shown that the literature on \emph{(de)cen\-tral\-iza\-tion} has exponentially increased in the past 70 years, with an author in 154 contributing to articles on the topic in 2021. 
\newtext{We have observed a diversification of research fields engaging in (de)centralization studies, starting from Governance as the most prolific field and gradually expanding to include various other disciplines.}
\deletetext{Through the analysis of the evolution of knowledge flows between clusters, we have revealed that, while initially the different fields had little \newtext{to no} communication with one another, they went on to increas cross-pollinations over time, especially within STEM. }
\newtext{Furthermore, analyzing the evolution of knowledge flows between clusters, we have revealed a gradual increase of influence between different fields.
Initially, the various fields operated in isolation, with minimal cross-disciplinary interaction.
However, as the literature developed, we observed a growing interconnection among these fields, with high knowledge flows especially within STEM subjects.}
Finally, we have shown how Governance has lost its leading role \newtext{in the \emph{(de)cen\-tral\-iza\-tion} literature} in favour of Blockchain\deletetext{,}\newtext{.}
\newtext{In fact, if Governance has remained mostly independent after the 1990s, without influencing or being influenced by other clusters, Blockchain} \deletetext{which }has been the most influential cluster for the last ten years\deletetext{ of the \emph{(de)cen\-tral\-iza\-tion} literature}, and has recently become the most productive one. 

\deletetext{Importantly, the framework we have developed for our analysis is general and may be used to analyse the history of any concept in the academic literature. }%
\newtext{A significant aspect of the framework we have developed is its versatility. The methodology can be used to examine the evolution of any scholarly term or concept within academic literature. For example, future work could use it to investigate the unfolding of such important topics as ``gender inequality" and ``artificial intelligence". Additionally, it can be utilized to explore the interplay between collective and independent innovation in the field of science.}
Our pipeline relies on two key methods, the multilayer hierarchical stochastic block model~\cite{peixoto_epj} and knowledge flows~\cite{knowledge_flow}.
On the one hand, we employ the first one to cluster documents and words in the dataset to identify different themes and topics, using information of both citations between papers and of the words used in each document. On the other hand, knowledge flows allow us to identify significant influences between clusters over time. 
With the present paper, we publicly release the pipeline code to allow other researchers to perform similar analyses on other concepts.

Our study presents some limitations which also represent directions for future work.
First, we only consider academic papers that directly mention the word \emph{(de)cen\-tral\-iza\-tion} or one of its variants (e.g. ``centralised", ``centralizing", etc.). A broadened analysis could also include all articles cited by these papers, in order to further understand the roots of this topic in the different fields.
Moreover, we have limited the semantic information to the \newtextnew{words of }document titles. 
Future studies could build on state of the art large scale language models and Natural Language Processing techniques to extract more information from the articles text (i.e., abstract and/or full text) and offer more detailed insights of their content.
\newtextnew{
For example, a set of keywords could be used instead of the plain text of the title~\cite{campos2020yake}, so as to better characterize papers and disambiguate terms that could have different meanings in different contexts.
In fact, the stochastic block model assigns one block to each node of the network~\cite{bouveyron2018stochastic,peixoto_epj}, thus indirectly assigning the main meaning of a word to a certain topic, disregarding other nuances of the word. A possible way to take this into account is to consider mixed-membership stochastic block models~\cite{airoldi2008mixed,zhu2013scalable}, where each node is assigned to a distribution or mixture of categories.
% Further notice that the presence our results are still robust to this issue, thanks to the use of multiple sources of information in our method, i.e., text and citations.
}

Finally, our methodology is able to identify direct flows of knowledge between two fields but misses less straightforward chains of interaction\deletetextnew{ (e.g., field $a$ influencing field $b$, which in turn influences field $c$, hence providing a possible indirect impact of $a$ to $c$).}\newtextnew{. For example, a field $a$ could indirectly influence field $b$ if there is a direct knowledge flow between field $a$ and another field $c$, which in turn influences field $b$.
} 
The inclusion of temporally and causally compatible higher order interactions (i.e., more than pairwise) is therefore an obvious route to improve on the current work.
\deletetextnew{Furthermore }\newtextnew{Moreover}\newtext{, it would be interesting to investigate the presence of }\newtextnew{citation }\newtext{biases or other spurious effects that have affected the evolution of science.}
For\deletetextnew{ example}\newtextnew{ instance}\newtext{, the rise of interdisciplinary knowledge flows could have been favoured by the presence of online easily-accessible papers.}
\newtextnew{Furthermore, 
even though the combination of clustering and knowledge flows are a powerful tool to detect statistically significant influence across fields and time, 
it is worth noticing that citations can be noisy, and are not always a definitive indicator of cross-pollination. 
We hence emphasize the need for complementary methods and approaches, such as qualitative historical analyses and expert interviews, to provide a more comprehensive understanding of the development of the concept of ``decentralization" in different disciplines.}
\newtext{We leave such research questions to future work.}

Overall, our \deletetext{work }\newtext{analysis} provides new insights in the origin and evolution of the ubiquitous concept of \emph{(de)cen\-tral\-iza\-tion}, shed\deletetext{s}\newtext{ding} light on the academic roots and influence of the blockchain technology\deletetext{, and offers a pipeline}\newtext{.}
\newtext{Additionally, our pipeline can be used } to analyse quantitatively any other concept in the academic literature\newtext{, and can be easily combined with other text and network analysis tools}. 
We therefore anticipate that our results will be of interest to researchers working in a vast array of disciplines. 

\section*{Availability of data and material}

All the code used for the pipeline presented in this paper can be freely accessed and used through the Github repository available at \url{https://github.com/alberto-bracci/decentralization}. The data used in this work can be obtained applying the pipeline to the open-access S2AG corpus, available at \url{https://www.semanticscholar.org/}.
All computational work has been conducted on high memory nodes of the HPC cluster of Queen Mary University of London~\cite{apocrita2017qmul}. 

\section*{Competing interests}
The authors declare no conflict of interest.

\section*{Funding}
This research received no specific grant from any funding agency in the public, commercial, or not-for-profit sectors. 

\section*{Author's contributions}
G.D.B, A.Br., N.P., V.L. and A.Ba. designed the study. G.D.B. and A.Br. carried out the data collection. G.D.B. and A.Br. performed the measurements. G.D.B, A.Br., N.P., V.L. and A.Ba. analysed the data, discussed the results, and contributed to the final manuscript.

\section*{Acknowledgements}
The authors wish to thank the anonymous referees for their valuable input on an earlier version of the manuscript.

\section*{List of abbreviations}
\textit{COVID-19}: COronaVIrus Disease of 2019.
\textit{NFT}: Non-Fungible Token.
\textit{STEM}: Science, Technology, Engineering, Mathematics.
\textit{S2AG}: Semantic Scholar Academic Graph.
\textit{MAG}: Microsoft Academic Graph.
\textit{hSBM}: hierarchical Stochastic Block Model.
\textit{SM}: Supplementary Material.

\section*{Additional Files}
  \subsection*{Supplementary Material}
    The interested reader can find more analyses in the Supplementary Material about the dataset (Section S1) and the hierarchical clustering method (Section S2), as well as other results about the importance of Governance and Blockchain in the history of (de)centralization (Section S3).

\clearpage

\onehalfspacing
\begin{Large}{\begin{center}
{\bf{Supplementary Information for 
``The Concept of Decentralization Through Time and Disciplines: A Quantitative Exploration"
}}
\end{center} 
 }
\end{Large}
\thispagestyle{empty}
\tableofcontents
\clearpage

\renewcommand{\thefigure}{S\arabic{figure}}
\setcounter{figure}{0}
\renewcommand{\thetable}{S\arabic{table}}
\setcounter{table}{0}
\renewcommand{\thesection}{S\arabic{section}}
\setcounter{section}{0}

%%%%%%%%%%%%%%%%%%%%%%%%%%%%%%%%%%%%%%%%%%
% Dataset info
%%%%%%%%%%%%%%%%%%%%%%%%%%%%%%%%%%%%%%%%%%

\section{Dataset}

In this section we report additional information on the S2AG corpus and the \emph{(de)centralization} dataset to complement the description in the article. 
\newtext{The dataset has been obtained by filtering the S2AG corpus, considering only papers that include words with the string ``centrali" in the title or in the abstract. To analyze which keywords are actually found, we count the number of occurrences of all the words that include the string ``centrali" in the dataset. In Table~\ref{table:no_papers_centrali} we show the count of all keywords appearing more than 1\,000 times. The keywords with the highest frequencies are ``(de)centralized" and ``(de)centralization". In order of frequency, the next keywords are alternative british versions of these terms. All other words are either less rare English words with the same root, such as ``centralizing", and translation of such words in other languages.}
\begin{table}[!b]
\begin{center}
\begin{tabular}{lr}
\toprule
          keyword &  count \\
\midrule
      centralized & 193\,513 \\
    decentralized & 175\,397 \\
 decentralization &  83\,304 \\
   centralization &  34\,328 \\
 decentralisation &  24\,231 \\
      centralised &  23\,572 \\
    decentralised &  23\,190 \\
   centralisation &   9\,424 \\
     centralizing &   6\,519 \\
descentralizacion &   6\,059 \\
      centralizer &   4\,969 \\
        centralis &   4\,850 \\
       centralize &   4\,647 \\
  descentralizado &   4\,295 \\
   decentralizing &   4\,126 \\
descentralización &   4\,046 \\
 descentralizacao &   3\,935 \\
     decentralize &   3\,591 \\
       centralism &   3\,546 \\
\bottomrule
\end{tabular}
\begin{tabular}{lr}
\toprule
         keyword &  count \\
\midrule
décentralisation &   2\,860 \\
      centralise &   2\,676 \\
 descentralizada &   2\,669 \\
    centralizada &   2\,577 \\
    centralizers &   2\,544 \\
    centralizado &   2\,369 \\
      centrality &   2\,061 \\
     centralisee &   2\,023 \\
   decentralisee &   1\,888 \\
descentralização &   1\,754 \\
    decentralise &   1\,571 \\
   centralizacao &   1\,496 \\
  centralizacion &   1\,472 \\
     centralizes &   1\,460 \\
descentralizados &   1\,422 \\
      centralist &   1\,319 \\
    centralising &   1\,210 \\
     centralismo &   1\,013 \\
\bottomrule
     & \\
\end{tabular}
\end{center}
\caption{\newtext{Number of occurrences of the keywords including the string ``centrali" in the dataset.}
}
\label{table:no_papers_centrali}
\end{table}

\begin{figure}[!htb]
    \centering
    \includegraphics[width=1.\textwidth]{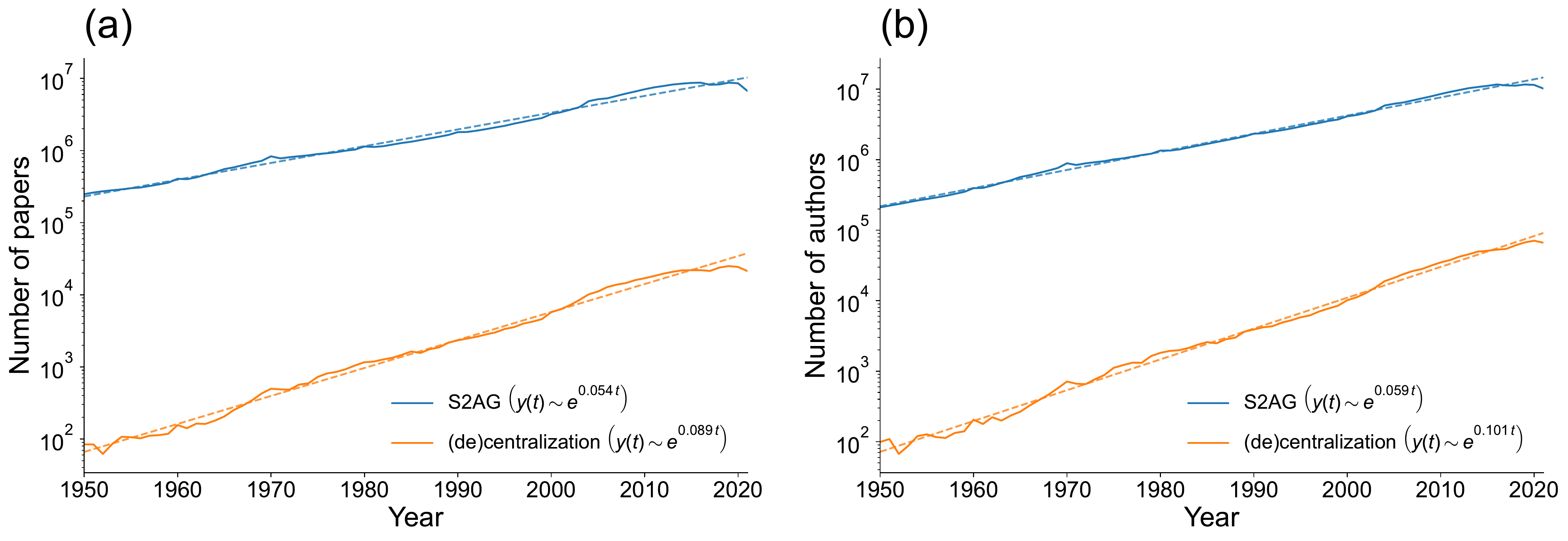}
    \caption{Comparison of the increase in number of papers \textbf{(a)} and authors \textbf{(b)} between the whole S2AG corpus and the \emph{(de)centralization} dataset. Exponential fits are shown as dashed line, with the fitted values between brackets in the legend.}
    \label{fig:num_papers_authors}
\end{figure}

\deletetext{I}\newtext{Moreover, i}n Table~\ref{table:no_papers_attributes} we report the number of publications in our dataset containing information about each attribute. 
In Table~\ref{table:no_papers_fieldsOfStudy} we report information on the publications' fields of study, as originally reported in the S2AG metadata. 
Each item can be classified in one or more fields. In the left table, we report all single fields or pairs that have more than 1000 documents in our dataset (papers can be classified in more than two fields, but the fields after the second are here discarded). 
In the right table we instead see how many papers have each of the 19 fields as first field in their classification, showing the predominance of Computer Science, and STEM subjects in general, in our dataset.
Finally, in Fig.~\ref{fig:num_papers_authors} we show the raw number of papers and authors each year for the full S2AG corpus, compared with the \emph{(de)centralization} dataset, showing how \emph{(de)centralization} has a faster exponential growth than the general academic literature as indexed by semantic scholar, as the different exponent of the fitted exponential functions clearly show. 

\begin{table}[!ht]
\begin{center}
\begin{tabular}{lr}
\toprule
Attribute & \#Publications \\
\midrule
\textit{title}              & 425\,144 \\
\textit{paperAbstract}      & 396\,201 \\
\textit{authors}            & 421\,611 \\
\textit{year}               & 423\,431 \\
\textit{inCitations} or 
\textit{outCitations}       & 305\,639 \\
\textit{fieldsOfStudy}      & 377\,720 \\
\textit{doi}                & 253\,464 \\
\textit{venue}              & 143\,802 \\
\textit{journalName}        & 234\,888 \\
\bottomrule
\end{tabular}
\end{center}
\caption{Number of papers (right column) in the \emph{(de)centralization} dataset that contain information on the listed attributes (left column). The attribute \textit{``inCitations"} and \textit{``outCitations"} correspond respectively to received citations and given references.
}
\label{table:no_papers_attributes}
\end{table}

\begin{table}[!ht]
\begin{center}
\begin{tabular}{lr}
\toprule
                  fieldsOfStudy &  \#Publications \\
\midrule
            (Computer Science,) &      99\,230 \\
          (Political Science,) &      49\,987 \\
                 (Engineering,) &      37\,469 \\
                    (Medicine,) &      29\,223 \\
                    (Business,) &      29\,134 \\
                  (Economics,) &      26\,716 \\
                  (Sociology,) &      15\,869 \\
                  (Geography,) &      13\,211 \\
                 (Mathematics,) &       8\,192 \\
                         (Art,) &       6\,474 \\
                  (Philosophy,) &       5\,767 \\
      (Environmental Science,) &       5\,708 \\
(Engineering, Computer Science) &       4\,482 \\
                     (History,) &       3\,941 \\
                  (Psychology,) &       3\,702 \\
                     (Physics,) &       3\,406 \\
(Mathematics, Computer Science) &       2\,633 \\
(Computer Science, Mathematics) &       2\,599 \\
(Computer Science, Engineering) &       2\,362 \\
                     (Biology,) &       2\,198 \\
            (Biology, Medicine) &       2\,025 \\
          (Materials Science,) &       2\,007 \\
  (Computer Science, Medicine) &       1\,825 \\
                  (Chemistry,) &       1\,394 \\
                     (Geology,) &       1\,194 \\
          (Business, Medicine) &       1\,167 \\
  (Business, Computer Science) &       1\,070 \\
\bottomrule
\end{tabular}
\qquad
\begin{tabular}{lr}
\toprule
            1st\_field &  \#Publications \\
\midrule
     Computer Science &     121\,600 \\
    Political Science &      58\,140 \\
          Engineering &      48\,458 \\
             Business &      35\,955 \\
             Medicine &      35\,639 \\
            Economics &      31\,842 \\
            Sociology &      18\,713 \\
            Geography &      15\,746 \\
          Mathematics &      12\,563 \\
                  Art &       7\,521 \\
Environmental Science &       6\,974 \\
          Philosophy &       6\,638 \\
          Psychology &       5\,115 \\
              Biology &       4\,944 \\
              History &       4\,587 \\
              Physics &       4\,355 \\
            Chemistry &       2\,484 \\
    Materials Science &       2\,477 \\
              Geology &       1\,393 \\
\bottomrule
\end{tabular}
\end{center}
\caption{Number of papers in decreasing order in the dataset that have have been categorized with the respective tuple of fields of study (left table) and with the respective first field (right table). Here we have listed only the fields of study with more than 1000 papers in the dataset. Tuples with more than two fields of study have been reduced considering only the first two. 
}
\label{table:no_papers_fieldsOfStudy}
\end{table}

%%%%%%%%%%%%%%%%%%%%%%%%%%%%%%%%%%%%%%%%%%
% hSBM
%%%%%%%%%%%%%%%%%%%%%%%%%%%%%%%%%%%%%%%%%%

\clearpage
\section{Hierarchical clustering}

\subsection{Partition overlap with the consensus partition}

\begin{figure}[!b]
    \centering
    \includegraphics[width=\textwidth]{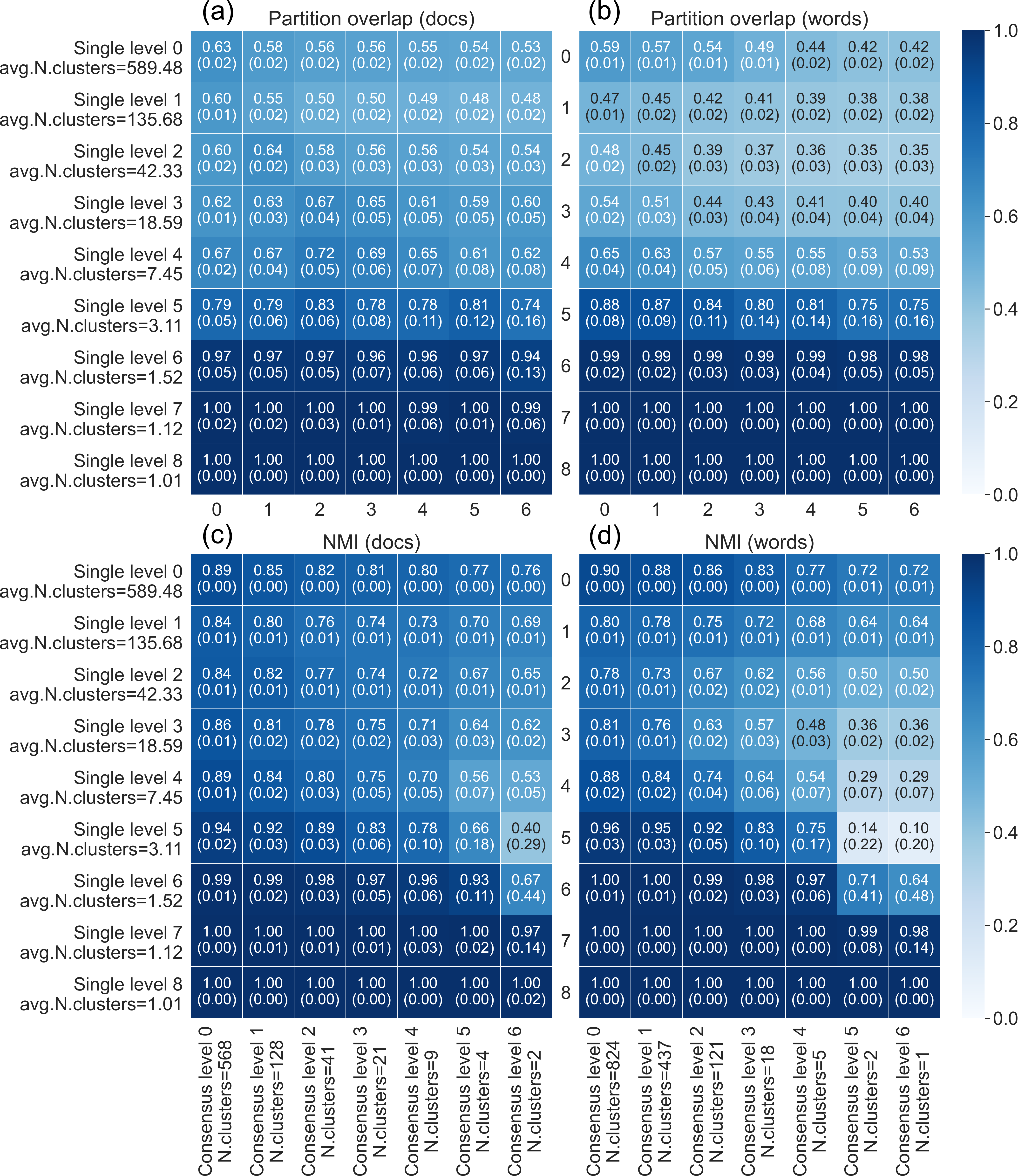}
    \caption{\textbf{Overlap between single iterations of the hSBM with the consensus partitions.}
    Maximum overlap \textbf{(a-b)} and Normalized Mutual Information \textbf{(c-d)} between the document \textbf{(a,c)} and word \textbf{(b,d)} partitions obtained from the different single iterations of the multilayer hSBM versus the consensus partition. Such consensus partitionˆis obtained by maximizing the sum of overlaps with all the partitions. 
    The value of a cell in row $i$ and column $j$ in the panel refers to average (and standard deviation within parentheses) for the single partitions at level $i$ and the consensus partition at level $j$.
    }
    \label{fig:doc_consensus_partition_nmi}
\end{figure}

\newtext{
In the pipeline of our analysis, an important step is the use of the \textit{multilayer hierarchical stochastic blockmodel} (or hSBM). 
We run 100 iterations of the algorithm and derive the consensus partition to increase robustness and generalizability of our results. 
This approach helps to account for the inherent stochasticity and variability in the algorithm, ensuring that the identified partition is not an artifact of a single run but a consistent representation of the data.
}

\newtext{
In Fig.~\ref{fig:doc_consensus_partition_nmi}, we present the maximum overlap and the Normalized Mutual Information (NMI) between the document and word partitions obtained from different single iterations of the multilayer hSBM, compared to the consensus partition. The panels (a) and (b) represent the maximum overlap, while panels (c) and (d) show the NMI. Moreover, panels (a) and (c) represent the document clusters, while panels (b) and (d) refer to the word clusters. 
}
\newtext{
The values in each cell $(i,j)$ of the panel indicate the average overlap (with standard deviation among parentheses) for the single partitions at level $i$ and the consensus partition at level $j$. 
This provides a comprehensive view of the agreement between individual iterations and the consensus partition. 
Moreover, the number of clusters at the various levels is displayed in the figure.
}

\newtext{
One notable finding is the high average overlap between the single partitions and the consensus partition, particularly at specific levels. This indicates a strong degree of agreement among the different iterations and the consensus, suggesting robustness and reliability in the identified partitions. 
Less agreement is instead found in the low levels, which have a higher level of granularity. For example, at level 0 the partitions have more than 500 clusters.
}

\newtext{
Furthermore, the NMI values provide insights into the information shared between the individual partitions and the consensus one. Higher NMI values indicate a greater level of similarity and shared information. The observed NMI values demonstrate a significant degree of agreement between the single partitions and the consensus partition, further supporting the reliability of the identified partitions.
}

\newtext{
Overall, these results highlight the effectiveness of the consensus partition in capturing the shared structure across multiple iterations of the hSBM. The high levels of overlap and NMI indicate a consistent and reliable identification of partitions, contributing to a more accurate understanding of the underlying patterns and structures within the analyzed data.
}

\newtext{
It should be noted that these results are specific to the current dataset and methodology employed. 
Further investigations and comparisons with alternative approaches are necessary to validate and generalize these findings. 
Nonetheless, the observed agreement between the single iterations and the consensus partition provides a promising avenue for future research in network analysis and partitioning algorithms.
}

\subsection{Keywords annotation}

\begin{figure}[!ht]
    \centering
    \includegraphics[width=\textwidth]{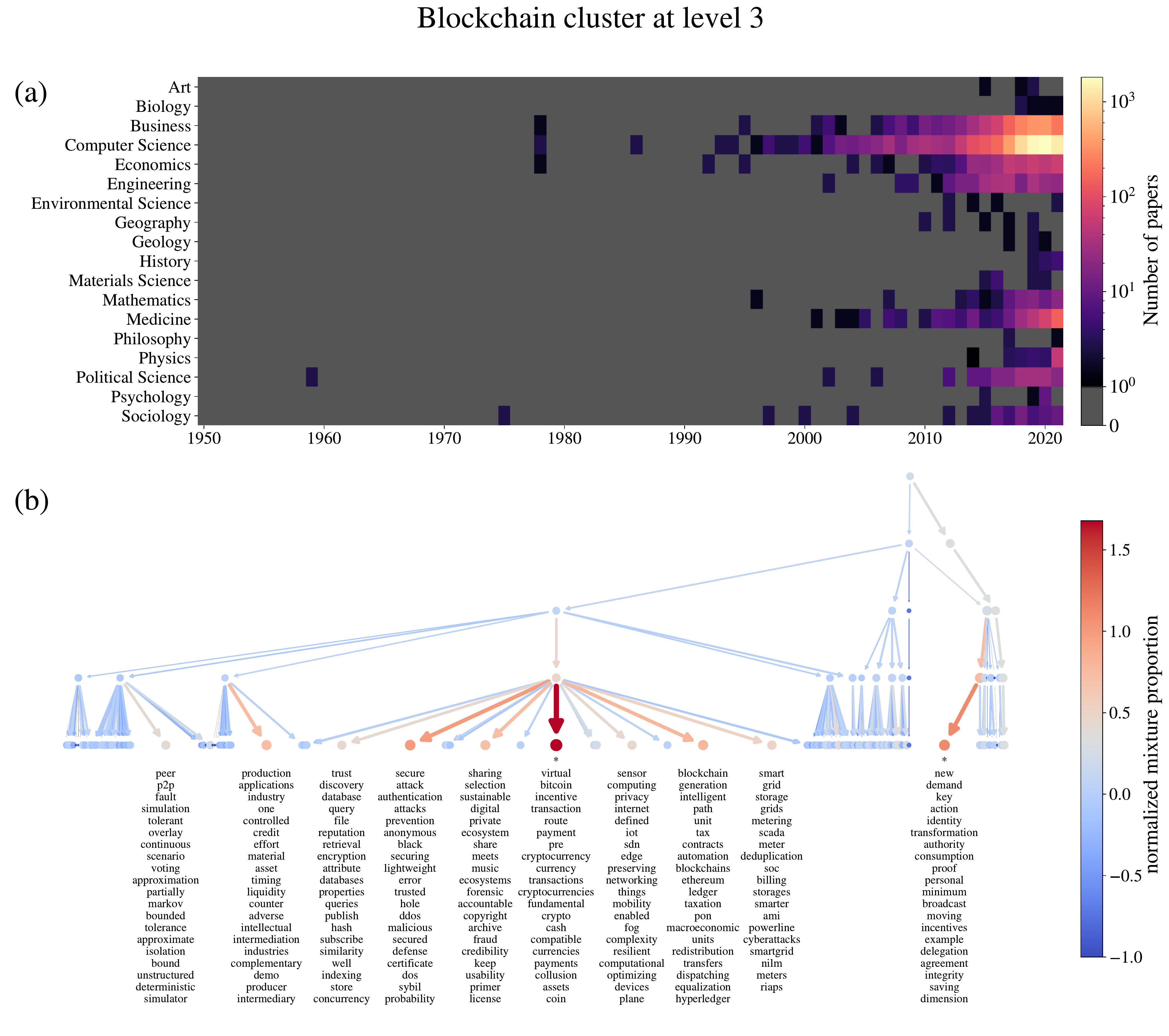}
    \caption{\textbf{Keywords annotation for Blockchain.}
    \textbf{(a)} Number of papers in time in each field of study for the Blockchain cluster.
    \textbf{(b)} Hierarchical topic tree, highlighting the top 20 words of the 10 most significant topics in the Blockchain cluster according to the normalized mixture proportion (in color).
    This information is used to aid the keywords annotation manual procedure.
    }
    \label{fig:annotation_blockchain}
\end{figure}

\begin{figure}[!ht]
    \centering
    \includegraphics[width=\textwidth]{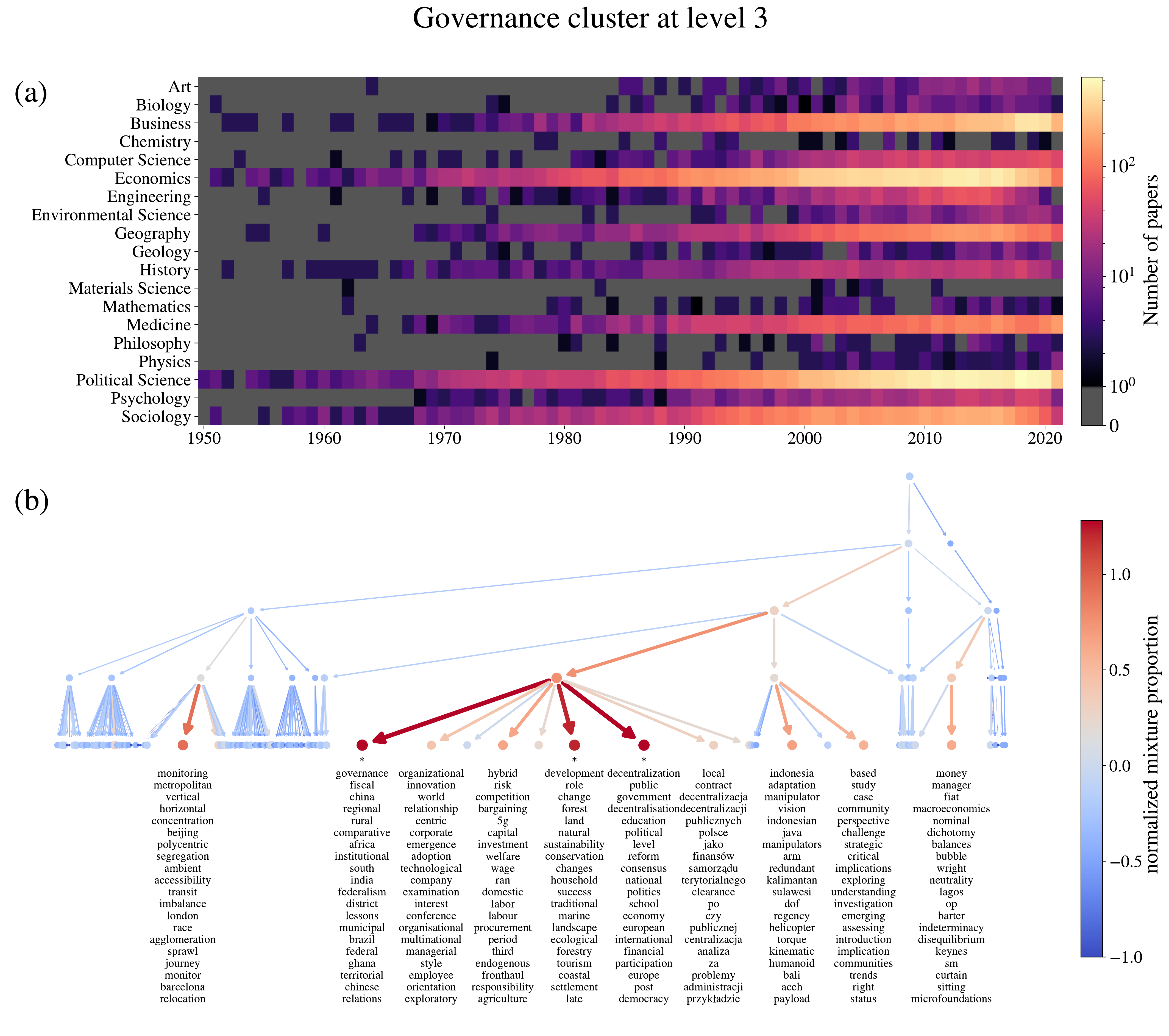}
    \caption{\textbf{Keywords annotation for Governance.}
    \textbf{(a)} Number of papers in time in each field of study for the Governance cluster.
    \textbf{(b)} Hierarchical topic tree, highlighting the top 20 words of the 10 most significant topics in the Governance cluster according to the normalized mixture proportion (in color).
    This information is used to aid the keywords annotation manual procedure.
    }
    \label{fig:annotation_governance}
\end{figure}

As described in the main text, the hSBM algorithm produces a hierarchical clustering of both articles and words (i.e. topics).
Here we give more details on the procedure we used to manually assign keywords to represent the different document clusters.
First, for each cluster at level 4, we manually inspect the most frequent words in the publications' titles. 
We then look at the most significant topics represented in the articles, as quantified by the normalized mixture proportion~\cite{peixoto_epj}. 
For instance, as we can see depicted in Fig.~\ref{fig:annotation_blockchain}(b) for the Blockchain cluster, or in Fig.~\ref{fig:annotation_governance}(b) for the Governance cluster, we can visualize the hierarchical tree of topics, highlighting the top 10 most significant ones. We have also reported in the tree the significance level ($> 1$) with the * symbol. 
Moreover, for each of these topic, the top 20 most significant words are printed.
We also look at the fields of study represented in each cluster in time, as shown in Fig.~\ref{fig:annotation_blockchain}(a) for Blockchain, or in Fig.~\ref{fig:annotation_governance}(a) for Governance. 
Finally, we look at the most important articles in the cluster, as represented by different metrics: overall number of citations, number of citations within the \emph{(de)centralization} dataset, highest knowledge flows towards other clusters, and most central publications according to different network centrality measures (degree, pagerank, betweenness, closeness, katz).

For clarity, we have then chosen one representative keyword for each cluster at the 3\textsuperscript{rd} level. In Table~\ref{table:no_papers_clusters}, we show the results of the keyword annotation procedure. 
Starting from the leftmost column, we show each cluster at level 3 with its representative keyword and the number of publications in it, to then show the annotated keywords in each cluster at level 4 (and the number of publications) present in its hierarchical branch in separate rows.
Clusters with less than 500 documents have been disregarded.

\begin{table}[!ht]
\begin{center}
\begin{tabular}{llrlr}
\toprule
{} &                  keyword &  count &                                       sub-keywords &  count \\
\midrule
\multirow[t]{3}{*}{0} & \multirow[t]{3}{*}{Robot swarms} & \multirow[t]{3}{*}{5\,843}&routing, allocation, congestion &        909\\
    &&&robot swarms, supply chain &       4\,120 \\
    &&&planning, game theory &        693\\ 
\multirow[t]{3}{*}{1} & \multirow[t]{3}{*}{Cybersecurity} & \multirow[t]{3}{*}{18\,437} &cybersecurity, peer2peer, fault tolearance, attacks &      15\,189 \\
    &&&cloud, security &       3\,248 \\ 
\multirow[t]{3}{*}{2} & \multirow[t]{3}{*}{Smart grids} & \multirow[t]{3}{*}{867} &smart grids, energy trading, blockchain &        867 \\ 
\multirow[t]{3}{*}{3} & \multirow[t]{3}{*}{Blockchain} & \multirow[t]{3}{*}{9\,637} &blockchain, cryptocurrency, ethereum, bitcoin &       9\,634 \\ 
\multirow[t]{3}{*}{4} & \multirow[t]{3}{*}{Federated learning} & \multirow[t]{3}{*}{2\,043} &federated learning, deep learning, adversarial networks &       1\,565 \\ 
    \hline
\multirow[t]{3}{*}{5} & \multirow[t]{3}{*}{Optimization} & \multirow[t]{3}{*}{5\,878} &control theory, equilibrium &        926 \\
    &&&statistical learning, optimization, detection &       3\,065 \\
    &&&heterogeneity, links, web, health &       1\,887 \\ 
\multirow[t]{3}{*}{6} & \multirow[t]{3}{*}{Control theory} & \multirow[t]{3}{*}{19\,758} &control theory, nonlinear dynamics &       8\,634 \\
    &&&navigation, flocking, formation &       9\,661 \\
    &&&discrete-event systems, decision making &       1\,463 \\ 
\multirow[t]{3}{*}{7} & \multirow[t]{3}{*}{Electricity} & \multirow[t]{3}{*}{8\,090} &electricity, grids &       8\,090 \\ 
    \hline
\multirow[t]{3}{*}{8} & \multirow[t]{3}{*}{Investments} & \multirow[t]{3}{*}{8\,394} &networks, connectivity, synchronization, routing, topology &       3\,058 \\
    &&&investments, money, risk, algorithm &       5\,336 \\ 
\multirow[t]{3}{*}{9} & \multirow[t]{3}{*}{Edge-computing} & \multirow[t]{3}{*}{2\,970} &edge-computing, cloud-computing &       2\,970 \\ 
    \hline
\multirow[t]{3}{*}{10} & \multirow[t]{3}{*}{Telecommunication} & \multirow[t]{3}{*}{18\,804} &cellular networks, communication, radio &      12\,446 \\
    &&&wireless, localization &       1\,262 \\
    &&&routing, protocols, security &       5\,096 \\ 
    \hline
    \hline
\multirow[t]{3}{*}{11} & \multirow[t]{3}{*}{Wireless technologies} & \multirow[t]{3}{*}{2\,166} &spanish &       1\,251 \\
    &&&wireless, cybersecurity, allocation, decision making &        915 \\ 
    \hline
    \hline
\multirow[t]{3}{*}{12} & \multirow[t]{3}{*}{Governance} & \multirow[t]{3}{*}{40\,045} &governance, fiscal federalism, government, development &      19\,371 \\
    &&&natural resources, education, governance &      20\,674 \\ 
    \hline
\multirow[t]{3}{*}{13} & \multirow[t]{3}{*}{Environment} & \multirow[t]{3}{*}{15\,386} &routing &       1\,278 \\
    &&&renewable energy, wastewater, environment &       8\,348 \\
    &&&portoguese, latin america &       5\,760 \\ 
\multirow[t]{3}{*}{14} & \multirow[t]{3}{*}{Social network analysis} & \multirow[t]{3}{*}{10\,375} &supply chain, manufacturing, pricing &       3\,938 \\
    &&&social network analysis, organizations, firms &       6\,437 \\ 
\multirow[t]{3}{*}{15} & \multirow[t]{3}{*}{Health} & \multirow[t]{3}{*}{12\,343} &algebra, healthcare, symptoms &       6\,291 \\
    &&&hospitals, HIV, cancer, surgery &       6\,052 \\
\bottomrule
\end{tabular}
\end{center}
\caption{\textbf{Results of the keywords annotation procedure.}
Clusters at 3\textsuperscript{rd} level, with associated manually annotated keyword and number of publications, together with the same information for each of the clusters at hierarchical level 4 in its branch of the hierarchy. 
Single horizontal lines denote clusters at the 2\textsuperscript{nd} hiearchical level, while the double horizontal lines denote the three branches at the 1\textsuperscript{st} level.
Clusters with less than 500 documents have been disregarded.
}
\label{table:no_papers_clusters}
\end{table}

\clearpage

\subsection{Results of the hSBM}

\begin{figure}[!ht]
    \centering
    \includegraphics[width=\textwidth]{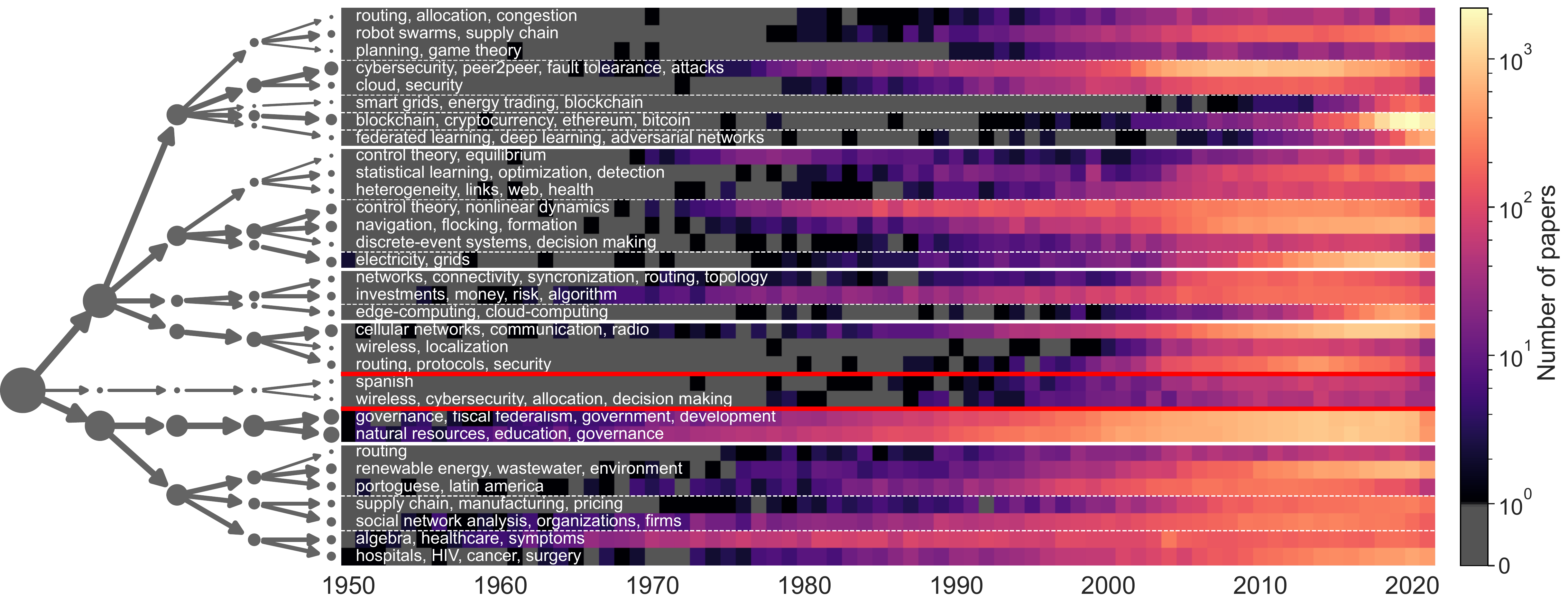}
    \caption{\textbf{Results of hSBM at the 4\textsuperscript{th} hierarchical level.}
    Heatmap of the number of papers in each cluster at level 4, together with the manually annotated keywords and the hierarchical tree resulting from the clustering algorithm.}
    \label{fig:results_methods_lev2}
\end{figure}

\begin{figure}[!ht]
    \centering
    \includegraphics[width=\textwidth]{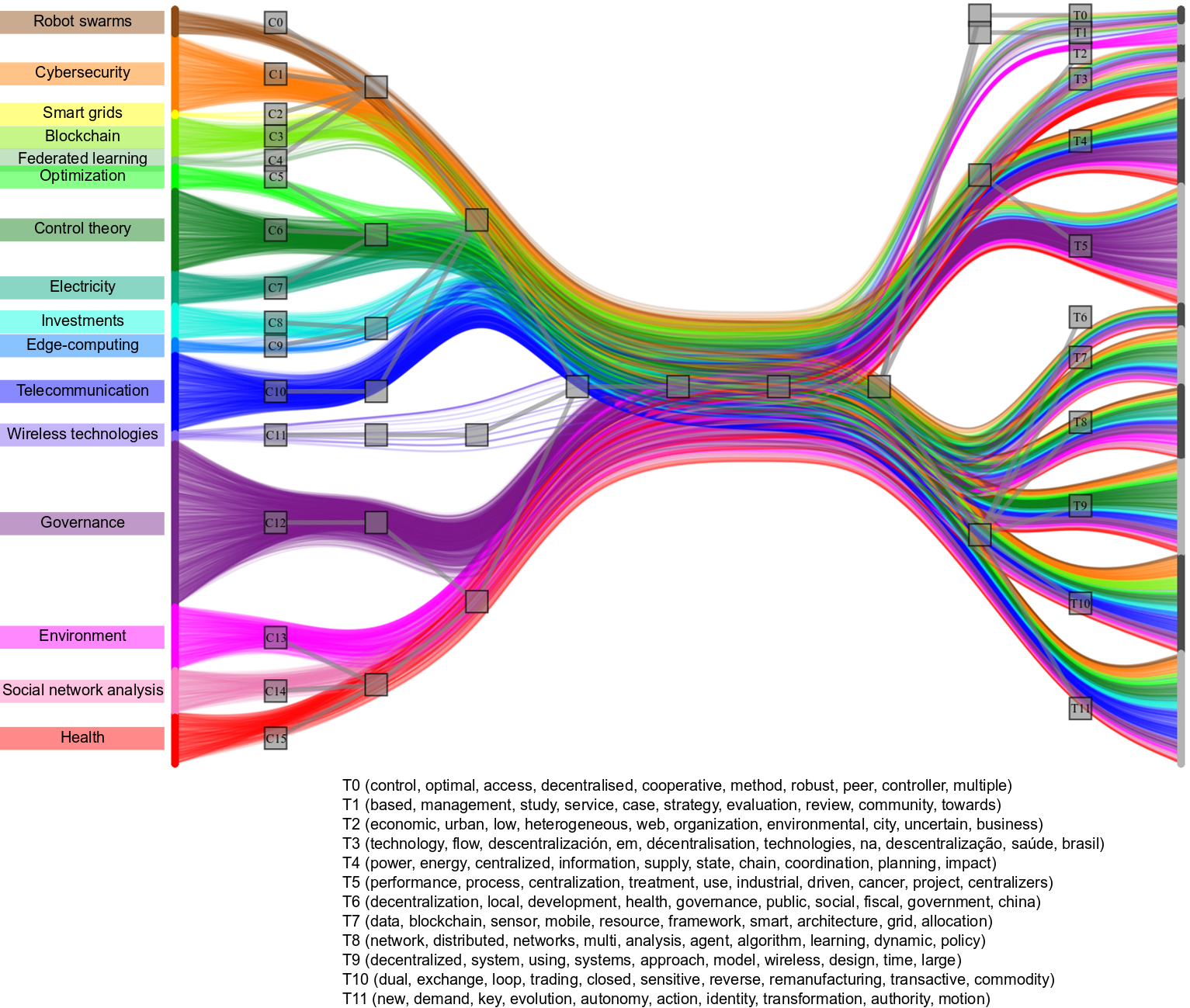}
    \caption{\textbf{Visualization of the results of the hSBM algorithm.}
    Bipartite network of documents and words, where nodes and links are colored based on the document cluster (at hierarchical level 3) they start from. Multiple instances of the same word are considered, one for each document the word is in. Word nodes within a topic at level 3 are ordered as the document clusters for visualization purposes.
    On top of the bipartite network, the hierarchy of topics and document clusters are represented on the word and document side respectively, shown as a network of square gray nodes.
    On the bottom of the figure, the legend with the 10 most frequent words of each topic is displayed.}
    \label{fig:bipartite_hsbm}
\end{figure}

In this section we report some additional visualizations of the results of the hSBM algorithm on the \emph{(de)centralization} dataset.
In Fig.~\ref{fig:results_methods_lev2} we show the number of papers in each cluster at hierarchical level 4, together with the hierarchical tree and the manually annotated keywords.
In Fig.~\ref{fig:bipartite_hsbm} we show the bipartite network of documents and words, together with the results of the hSBM algorithm and the manual annotation procedure. This figure shows how the various clusters are represented in terms of topic frequency.
Nodes on the left, indeed, represent documents, while those on the right title words. Notice that, to proper represent topic frequencies, multiple instances of the same word are considered, one for each document the word is in. 
Links are colored based on the doc cluster they start from.
On top of the bipartite network between docs and words, we show a tree where squares represent clusters/topics at various hierarchical levels as a result of the hSBM, starting from the common root to level 3.

\clearpage

\clearpage
\section{The importance of Governance and Blockchain in the history of \emph{(de)centralization}}

\begin{figure}[!ht]
    \centering
    \includegraphics[width=\textwidth]{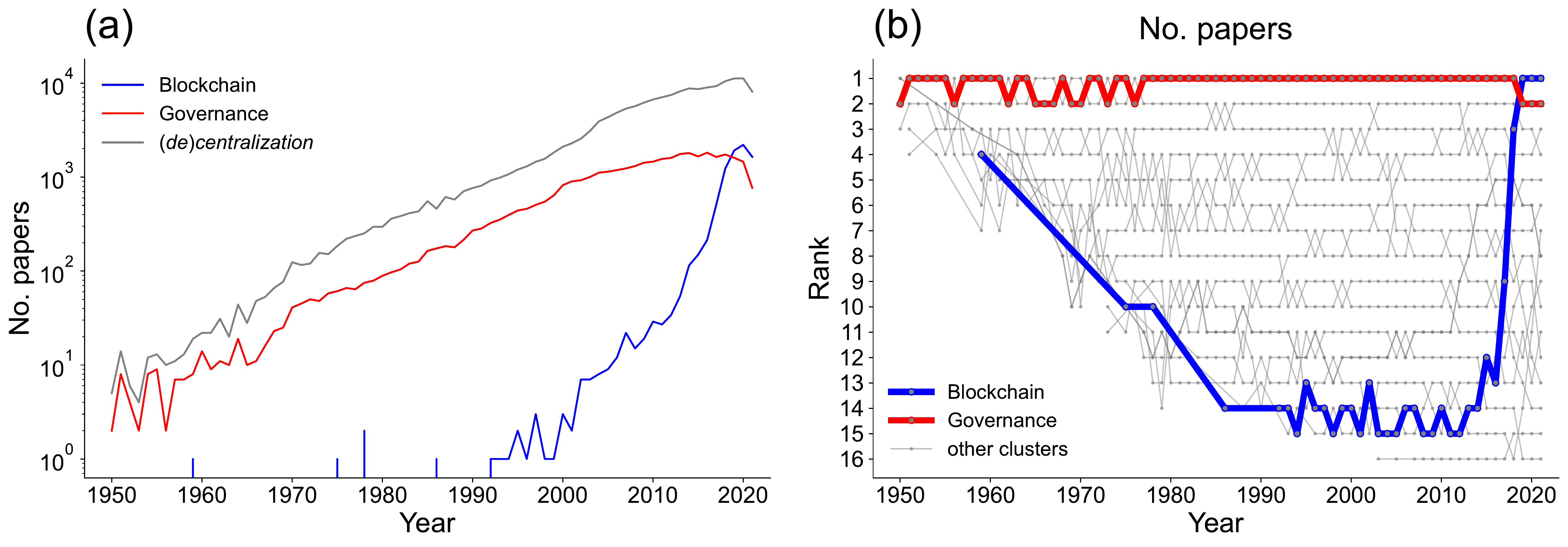}
    \caption{\textbf{Blockchain and Governance are the most productive clusters.}
    \textbf{(a)} Number of papers in time for the Blockchain cluster, the Governance cluster and the full \emph{(de)centralization} dataset.
    \textbf{(b)} Rank by number or papers in time for different clusters, highlighting Blockchain and Governance.
    Both plots show the central role of these two clusters in the literature on \emph{(de)centralization}, and how they have exchanged roles, with Blockchain becoming the most productive field in recent years.}
    \label{fig:fig3_no_papers}
\end{figure}

In this section we report some additional and complementary plots to show the importance of the Blockchain and Governance clusters in the academic literature on \emph{(de)centralization}.
In Fig.~\ref{fig:fig3_no_papers}(a) we show the number of papers in time for three groups of documents: Blockchain, Governance and the whole dataset. Moreover, in Fig.~\ref{fig:fig3_no_papers}(b) we plot the rank of each cluster in terms of yearly number of papers, highlighting Blockchain and Governance. Both figures clearly show how Governance has been the most productive cluster in the literature for a long time, only to be replaced by Blockchain in recent years.

\begin{figure}[!ht]
    \centering
    \includegraphics[width=\textwidth]{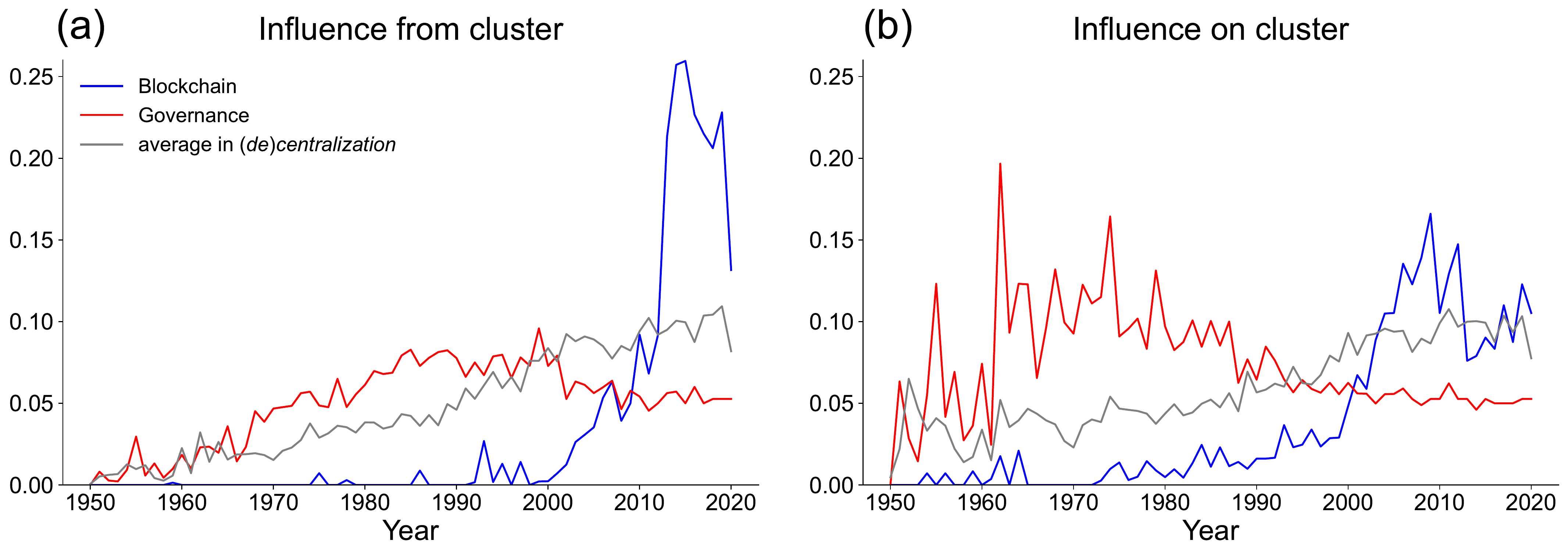}
    \caption{\textbf{Influence of Governance and Blockchain from and on other clusters in time.}
    Average knowledge flow towards other clusters in the future in time from a cluster \textbf{(a)} or to a cluster \textbf{(b)}, the latter being either Blockchain (in blue) or Governance (in red), while in gray the average over the \emph{(de)centralization} dataset is plotted.}
    \label{fig:fig3_kf}
\end{figure}

In Fig.~\ref{fig:fig3_kf} we look at knowledge flows, plotting their value instead of their rank (as done in Fig.~3 in the main text).
We plot the time evolution of $K_{a \to \bullet} (Y)$, where $a$ is Governance or Blockchain, in comparison to $K_{\bullet \to \bullet} (Y)$ in Fig.~\ref{fig:fig3_kf}(a). This shows the average influence that the cluster in a certain year has towards \deletetext{the other }\newtext{all }clusters in the future. We similarly compare the average influence of all other clusters on the future of these two clusters looking at $K_{\bullet \to a} (Y)$ in Fig.~\ref{fig:fig3_kf}(b).
We can see from these figures how Governance has been increasingly important in influencing other clusters until the 1980s, while since after the 1990s it has had a lower knowledge flow than the average among all clusters, despite being the first cluster in terms of number of papers in all this time.
The case of Blockchain is opposite: after 2013 it starts to have a much higher influence towards the other clusters compared to the average one over all clusters, while the other clusters after 2004 have had an average knowledge flow towards it. 

\begin{figure}[!ht]
    \centering
    \includegraphics[width=\textwidth]{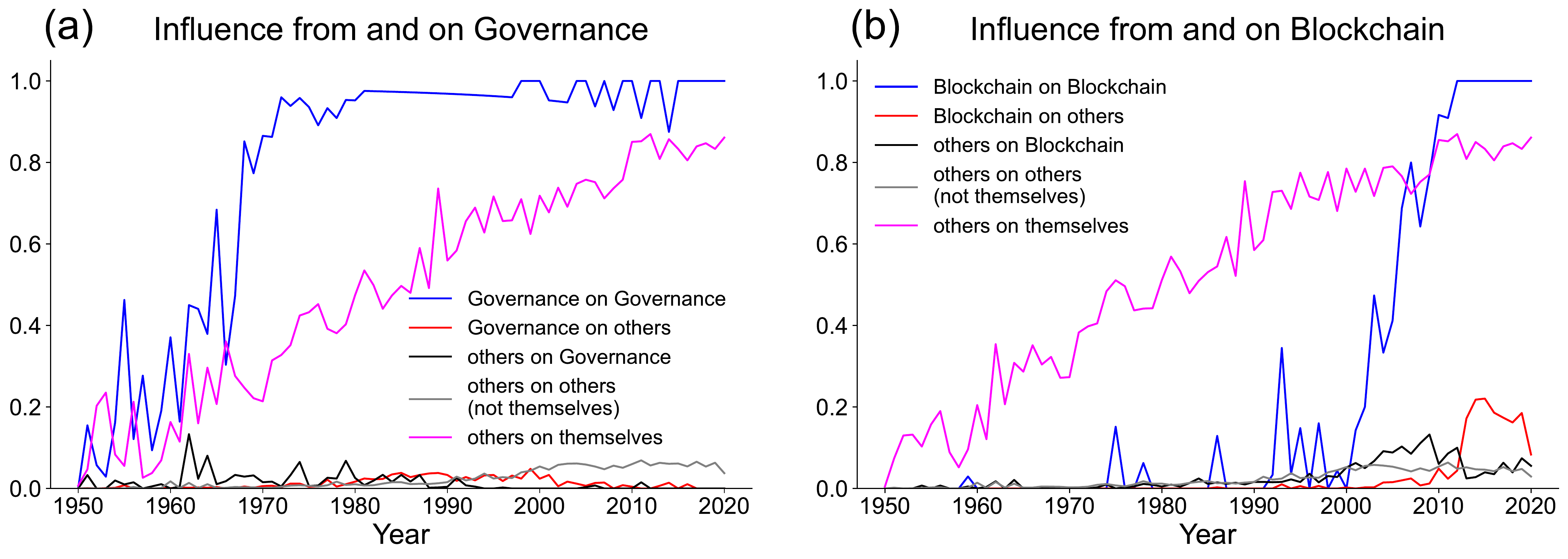}
    \caption{\newtext{\textbf{Influence of Governance and Blockchain from and on other clusters in time.}
    Average knowledge flows from or to Governance (\textbf{(a)}) or Blockchain (\textbf{(b)}) compared to the other clusters, distinguishing between the knowledge flow to the same cluster or to all other ones.}
    }
    \label{fig:fig3_kf_vs_themselves}
\end{figure}

\newtext{
In Fig.~\ref{fig:fig3_kf_vs_themselves}, we show the average knowledge flow of a cluster $a$ to the future, while also distinguishing the influence on the cluster itself. Fig.~\ref{fig:fig3_kf_vs_themselves}(a) focuses on the comparison of knowledge flows between Governance and other clusters, while Fig.~\ref{fig:fig3_kf_vs_themselves}(b) repeats the analysis on Blockchain.
Starting from the 1970s, we observe a substantial knowledge flow within the Governance cluster towards future Governance papers (blue line), consistently close to 1. In comparison, other clusters have had a lower impact on their future papers during this period (magenta line), with a more noticeable and increasing influence in recent years.
Examining the knowledge flow from Governance to other clusters (red line), we find that it has consistently been low, peaking in the 1990s. However, after the year 2000, the impact on other clusters has vanished. Conversely, all other clusters have exerted an increasing influence on all clusters excluding themselves (gray line), indicating their higher impact on other areas and their more efficient interdisciplinary communication.
Hence, while the number of papers in Governance has remained high, its impact has largely been confined to the same cluster after the 1990s. Conversely, the opposite trend is observed for other clusters, where their influence on other disciplines has progressively grown.
Furthermore, the influence of other clusters on Governance (black line) has been negligible since the 1990s, reinforcing the isolation of the Governance cluster.
In contrast, the Blockchain cluster, gaining significance in the 2000s (blue line in Fig.~\ref{fig:fig3_kf_vs_themselves}(b)), has received substantial knowledge flow from other clusters (black line), particularly during that period. This influx of knowledge has contributed to the emergence of blockchain technologies, which have subsequently made a significant impact on other clusters in recent years, as noticed by the high knowledge flow from Blockchain to other clusters (red line), thereby driving the recent scientific advancements in decentralization.
}

\clearpage
\subsection{Influence of Governance from and on other clusters}

Here we report an additional plot looking at the role of Governance with respect to the other clusters. 
In particular, we replicate Fig.~4 in the main text for Governance instead of Blockchain.
Given the results on the predominant role of Governance as an influence on other clusters before the year 2000, we look at the following three periods of time: from 1950 to 1980, from 1981 to 1990 and from 1991 to 2000, highlighting different interactions with other clusters both in terms of source and destination of knowledge flows. 
For instance, the Blockchain cluster is the second most influenced cluster by Governance in the 1990s, highlighting its role in the foundations of the field. 
However, the rankings are generally more stable than what observed in Fig.~4 in the main text for the case of the Blockchain cluster.

\begin{figure}[!ht]
    \centering
    \includegraphics[width=\textwidth]{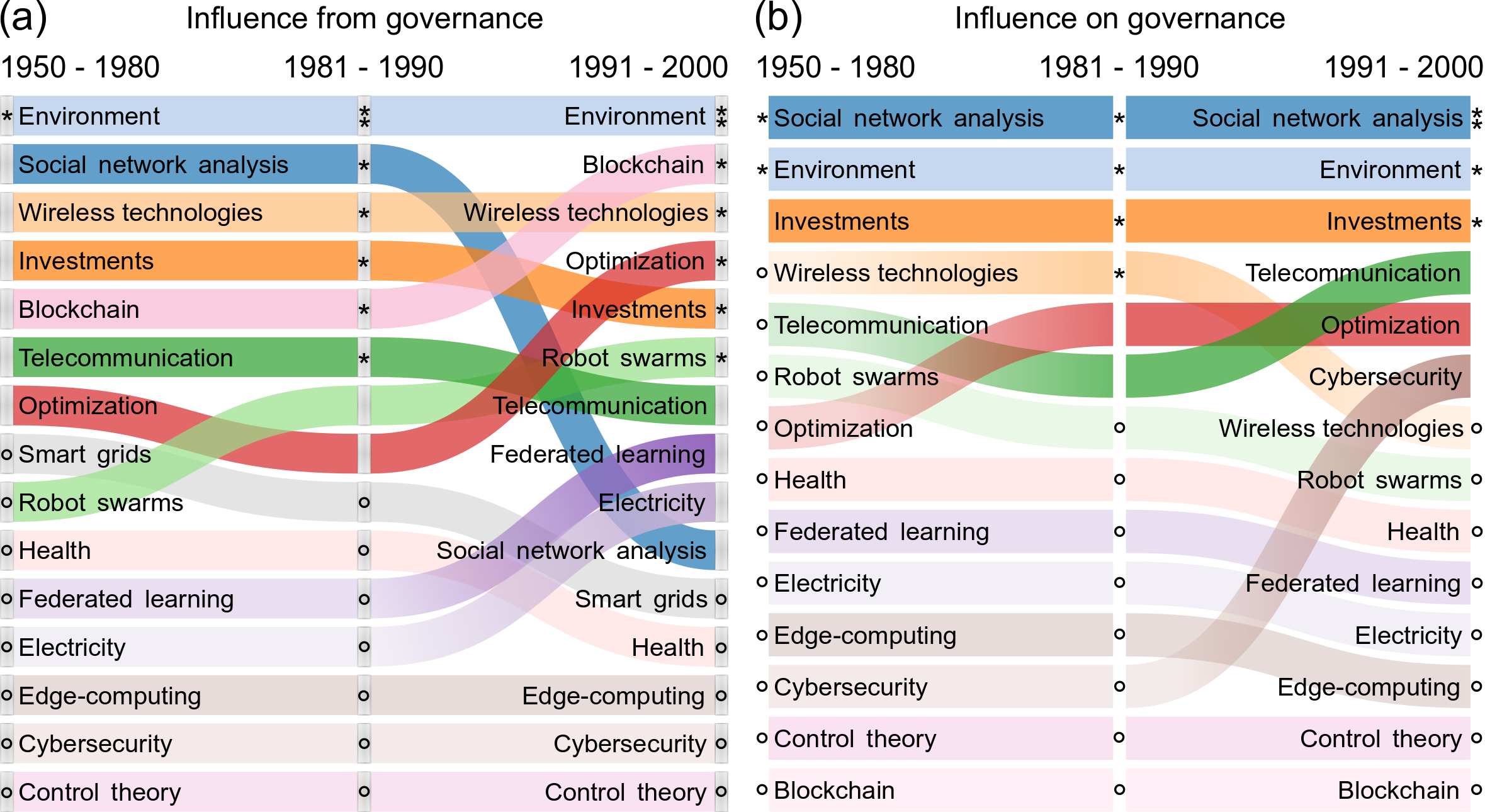}
    \caption{
    \textbf{The influence from/on Governance with regards to the rest of the \emph{(de)centralization} literature.} 
    \textbf{(a)} Change of the ranking of the clusters most influenced by the Governance literature between its early period (1950-1980), its middle period (1981-1990), and its late period (1991-2000), calculated using the average knowledge flows $K_{a \to b} (T)$, where $T$ is the selected period, and $a$ is fixed to Governance.
    \textbf{(b)} Change of the ranking of the most influential clusters on the Governance literature between its early period (1950-1980), its middle period (1981-1990), and its late period (1991-2000), calculated using the average knowledge flows $K_{a \to b} (T)$, where $T$ is the selected period, and $b$ is fixed to Governance.
    In both cases, if $K_{a \to b} (T) = 0$, we print a circle in the corresponding gray node and use a lighter color in the respective link. Moreover, we print a star when $0.01 < K_{a \to b} (T) \leq 0.1$, and two stars when $K_{a \to b} (T) > 0.1$.
    }
    \label{fig:fig4_governance}
\end{figure}

\end{document}